\documentclass[%
reprint,
superscriptaddress,
showpacs,
amsmath,amssymb,
prc,
floatfix, ]%
{revtex4-1}

\usepackage{color}
\usepackage{CJK}

\usepackage{graphicx}
\usepackage{dcolumn}
\usepackage{bm}
\usepackage[dvipdfmx,bookmarks=true,colorlinks,%
            citecolor=blue,linkcolor=blue,anchorcolor=blue,filecolor=blue,urlcolor=blue,%
           ]{hyperref}

\allowdisplaybreaks

\begin{document}

\begin{CJK*}{UTF8}{}

\title{Tetrahedral shapes of neutron-rich Zr isotopes from multidimensionally-constrained relativistic Hartree-Bogoliubov model}
\CJKfamily{gbsn}
\author{Jie Zhao (赵杰)}%
 \affiliation{CAS Key Laboratory of Theoretical Physics,
              Institute of Theoretical Physics, Chinese Academy of Sciences, Beijing 100190, China}
 \affiliation{Physics Department, Faculty of Science, University of Zagreb, Bijenicka 32,
              Zagreb 10000, Croatia}
\CJKfamily{gbsn}
\author{Bing-Nan Lu (吕炳楠)}%
 \affiliation{CAS Key Laboratory of Theoretical Physics,
              Institute of Theoretical Physics, Chinese Academy of Sciences, Beijing 100190, China}
 \affiliation{Institut f\"ur Kernphysik (IKP-3) and J\"ulich Center for Hadron Physics,
              Forschungszentrum J\"ulich, D-52425 J\"ulich, Germany}
\CJKfamily{gbsn}              
\author{En-Guang Zhao (赵恩广)}%
 \affiliation{CAS Key Laboratory of Theoretical Physics,
              Institute of Theoretical Physics, Chinese Academy of Sciences, Beijing 100190, China}
 \affiliation{Center of Theoretical Nuclear Physics, National Laboratory
              of Heavy Ion Accelerator, Lanzhou 730000, China}
\CJKfamily{gbsn}              
\author{Shan-Gui Zhou (周善贵)}%
 \email{sgzhou@itp.ac.cn}
 \affiliation{CAS Key Laboratory of Theoretical Physics,
              Institute of Theoretical Physics, Chinese Academy of Sciences, Beijing 100190, China}
 \affiliation{School of Physics, University of Chinese Academy of Sciences, Beijing 100049, China}
 \affiliation{Center of Theoretical Nuclear Physics, National Laboratory
              of Heavy Ion Accelerator, Lanzhou 730000, China}
 \affiliation{Synergetic Innovation Center for Quantum Effects and Application, 
              Hunan Normal University, Changsha, 410081, China}

\date{\today}

\begin{abstract}
We develop a multidimensionally constrained relativistic Hartree-Bogoliubov (MDC-RHB) model
in which the pairing correlations are taken into account by making the Bogoliubov transformation.
In this model, the nuclear shape is assumed to be invariant under the reversion of $x$ 
and $y$ axes; 
i.e., the intrinsic symmetry group is $V_4$ and all shape degrees of freedom 
$\beta_{\lambda\mu}$ with even $\mu$ are included self-consistently. 
The RHB equation is solved in an axially deformed harmonic oscillator basis.
A separable pairing force of finite range is adopted in the MDC-RHB model.
The potential energy curves of neutron-rich even-even Zr isotopes are calculated
with relativistic functionals DD-PC1 and PC-PK1
and possible tetrahedral shapes in the ground and isomeric states are investigated.
The ground state shape of $^{110}$Zr is predicted to be tetrahedral with 
both functionals and so is that of $^{112}$Zr with
the functional DD-PC1.
The tetrahedral ground states are caused by large energy gaps around $Z=40$ and $N=70$ 
when $\beta_{32}$ deformation is included.
Although the inclusion of the $\beta_{30}$ deformation can also reduce the energy 
around $\beta_{20}=0$ and lead to minima with pear-like shapes for nuclei around $^{110}$Zr, 
these minima are unstable due to their shallowness.   
\end{abstract}

\maketitle

\end{CJK*}

\section{Introduction~\label{sec:Introduction}}

The intrinsic shape of most known nuclei deviates from a sphere 
\cite{Bohr1969_Nucl_Structure,Bohr1975_Nucl_Structure}. 
The deformation of a nucleus can be described by the
parametrization of the nuclear surface with a multipole expansion,
\begin{equation}
 R ( \theta, \varphi ) = 
 R_0 \left[  1 + 
           \beta_{00} + 
           \sum_{\lambda=1}^{\infty} \sum_{\mu=-\lambda}^\lambda 
            \beta_{\lambda \mu}^* Y_{\lambda \mu} ( \theta, \varphi )    
     \right] ,
 \label{Eq:SurfaceDeformation}
\end{equation}
where $\beta_{\lambda \mu}$'s are deformation parameters. 
Various nuclear deformations characterized by $\beta_{\lambda \mu}$ with
different $\{\lambda, \mu\}$ are connected with many nuclear phenomena 
which have been predicted and some of them have also been observed. 

In recent decades, the triaxiality and reflection asymmetry in 
nuclear shapes have become more and more attractive because several interesting 
theoretical predictions related to these shapes have been confirmed experimentally. 
The static triaxial shapes may result in the wobbling motion 
\cite{Bohr1975_Nucl_Structure}
or chiral doublet bands \cite{Frauendorf1997_NPA617-131} in 
atomic nuclei; 
both phenomena
were observed in 2001 \cite{Odegard2001_PRL86-5866,Starosta2001_PRL86-971}.
In 2006, Meng et al. predicted that multiple chiral
doublet (M$\chi$D) bands may exist in one nucleus with different triaxial
configurations \cite{Meng2006_PRC73-037303}.
In recent years, M$\chi$D bands were indeed found in several nuclei 
\cite{Ayangeakaa2013_PRL110-172504,Lieder2014_PRL112-202502,%
Kuti2014_PRL113-032501,Tonev2014_PRL112-052501,Liu2016_PRL116-112501}.
The triaxiality may also play a role in the low-spin signature inversion
\cite{Bengtsson1984_NPA415-189,Liu1995_PRC52-2514,Liu1996_PRC54-719,Zhou1996_JPG22-415,%
Riedinger1997_PPNP38-251,Liu1998_PRC58-1849}.
Moreover, the development of triaxial shapes could lead to the termination of rotational bands 
in nuclei with $A \simeq 110$; see Ref.~\cite{Afanasjev1999_PR322-1} for a review. 
\label{page:termination}
The pear-like nuclear shape, characterized by $\beta_{30}$, was predicted 
to be very pronounced in many 
nuclei \cite{Butler1996_RMP68-349,Moeller2008_ADNDT94-758,Agbemava2016_PRC93-044304}.
\label{page:octupole}
Low-lying negative-parity levels which are connected with the ground state
band via strong $E1$ transitions in actinides and some
rare-earth nuclei are related to reflection asymmetric shapes 
with nonzero $\beta_{30}$ 
\cite{Shneidman2003_PRC67-014313,Shneidman2006_PRC74-034316,%
Wang2005_PRC72-024317,%
Yang2009_CPL26-082101,Robledo2011_PRC84-054302,Zhu2012_PRC85-014330,Zhu2012_NSC2012-348,%
Nomura2013_PRC88-021303R,Nomura2014_PRC89-024312,%
Nomura2015_PRC92-014312}.
The direct evidence for a static octupole deformation has been shown
experimentally in $^{224}$Ra \cite{Gaffney2013_Nature497-199}
and in $^{144}\mathrm{Ba}$ \cite{Bucher2016_PRL116-112503}.
\label{page:224Ra}

What new features can we expect if the nonaxial and reflection asymmetric
deformations are put together or combined?
It has been revealed in Ref.~\cite{Lu2012_PRC85-011301R} 
that both the triaxial and octupole shapes become important 
around the second saddle point in the potential energy surface (PES) of actinide 
nuclei and they lower the second fission barrier considerably.
Recently, Liu et al. observed for the first time octupole correlations between 
M$\chi$D bands in $^{78}$Br and
proposed that chirality-parity quartet bands may appear in a nucleus with 
both a static triaxial deformation ($\beta_{22}$) and an octupole deformation ($\beta_{30}$)
\cite{Liu2016_PRL116-112501}.

The nonaxiality and reflection asymmetry are combined intrinsically in deformations
characterized by $\beta_{\lambda\mu}$ with odd $\lambda$ and nonzero $\mu$.
Among such deformations, the $\beta_{32}$ deformation 
is of particular interest and has been investigated extensively
\cite{Hamamoto1991_ZPD21-163,Skalski1991_PRC43-140,Li1994_PRC49-R1250,%
Takami1998_PLB431-242,%
Yamagami2001_NPA693-579,%
Dudek2002_PRL88-252502,Dudek2006_PRL97-072501,%
Olbratowski2006_IJMPE15-333,%
Zberecki2006_PRC74-051302R,%
Dudek2010_JPG37-064032}.
A nucleus with a pure $\beta_{32}$ deformation, i.e.,
$\beta_{\lambda\mu}=0$ if $\lambda\ne3$ and $\mu\ne2$, 
has a tetrahedral shape with the symmetry group $T_d^D$. 
The study of single-particle structure of a nucleus with tetrahedral 
symmetry predicted large energy gaps at $Z (N)=16$, 20, 32, 40, 56--58, 70, and 90--94
and $N=112$ and 136/142 
\cite{Li1994_PRC49-R1250,Dudek2002_PRL88-252502,Dudek2007_IJMPE16-516,Dudek2003_APPB34-2491,
Heiss1999_PRC60-034303,Arita2014_PRC89-054308}.
These shell gaps are comparable to or even stronger than those at spherical shapes.
Thus, a nucleus with proton and/or neutron numbers equal to these values 
may have a static tetrahedral shape or strong tetrahedral correlations.

Various nuclei such as $^{80,108,110}$Zr, $^{160}$Yb, $^{156}$Gd, and $^{242}$Fm 
were predicted to have ground or isomeric states with tetrahedral shapes  
from the macroscopic-microscopic (MM) model 
\cite{Dudek2002_PRL88-252502,Dudek2007_IJMPE16-516,Dudek2006_PRL97-072501,%
Schunck2004_PRC69-061305R,Dudek2014_PS89-054007}
and
the Skyrme Hartree-Fock (SHF) model, 
the SHF plus BCS model, or the Skyrme Hartree-Fock-Bogoliubov (SHFB) model
\cite{Dudek2007_IJMPE16-516,Schunck2004_PRC69-061305R,Dudek2006_PRL97-072501,%
Yamagami2001_NPA693-579,Olbratowski2006_IJMPE15-333,Zberecki2009_PRC79-014319,%
Zberecki2006_PRC74-051302R,Takami1998_PLB431-242}.
The possible existence of tetrahedral symmetry in light nuclei such as 
$^{16}$O \cite{Bijker2014_PRL112-152501}
as well as superheavy nuclei \cite{Chen2013_NPR30-278,Chen2010_NPA834-378c} 
was also proposed.
Furthermore, the rotational properties of tetrahedral nuclei have 
been studied theoretically 
\cite{Gao2004_CPL21-806,Tagami2013_PRC87-054306,Tagami2015_JPG42-015106}
which would certainly be helpful for experimentalists to identify tetrahedral nuclei.

\label{page:expt1}
Several experiments were devoted to the study of tetrahedral shapes in 
$^{160}$Yb \cite{Bark2010_PRL104-022501},
$^{154,156}$Gd \cite{Bark2010_PRL104-022501,Jentschel2010_PRL104-222502,Doan2010_PRC82-067306},
$^{230,232}$U \cite{Ntshangase2010_PRC82-041305R}, 
and $^{108}$Zr \cite{Sumikama2011_PRL106-202501}.
For $^{160}$Yb and $^{154,156}$Gd, although the observed odd and even-spin members 
of the lowest energy negative-parity bands are similar to the proposed 
tetrahedral band of $^{156}$Gd, 
the deduced nonzero quadrupole moments existence of tetrahedral 
shapes in these nuclei.
The possibility of a tetrahedral shape for the negative-parity bands in 
$^{230,232}$U was excluded in Ref.~\cite{Ntshangase2010_PRC82-041305R}
by the observed similarity in the energies and electric dipole moments of 
those bands in the $N = 138$ and $N = 140$ isotones and 
the measured values for quadrupole moment $Q_2$ in $^{226}$Ra.
An isomeric state observed in $^{108}$Zr has been assigned a tetrahedral shape 
\cite{Sumikama2011_PRL106-202501}. 
However, this isomeric state can also been explained by assuming an axial shape
with the angular momentum projected shell model 
\cite{Liu2011_NPA858-11,Liu2015_SciChinaPMA58-112003}.

\label{page:expt2}
Besides the tetrahedral shape which corresponds to a pure $\beta_{32}$ deformation,
the study of nuclear shapes with a non-zero $\beta_{32}$ superposed on a sizable $\beta_{20}$
has become a hot topic too.
In recent years, nuclei with $Z\simeq 100$ have been studied extensively 
because such studies can not only reveal the structure for these nuclei 
but also give useful structure information for superheavy nuclei
\cite{Afanasjev2003_PRC67-024309,Leino2004_ARNPS54-175,Herzberg2006_Nature442-896,%
Herzberg2008_PPNP61-674,Zhang2011_PRC83-011304R,Zhang2012_PRC85-014324,Zhang2013_PRC87-054308,
Liu2014_PRC89-044304,
Shi2014_PRC89-034309,
Wang2014_CPC38-074101,
Afanasjev2014_PS89-054001,
Agbemava2015_PRC92-054310,
Afanasjev2015_JPG42-034002,
Li2016_SciChinaPMA59-672011}.
One example of such studies is the observation of very low-lying $2^-$ bands 
in several even-$Z$ $N = 150$ isotones \cite{Robinson2008_PRC78-034308}.
This low-lying feature was argued by Chen et al. to be
a fingerprint of sizable $\beta_{32}$ in these well deformed nuclei
\cite{Chen2008_PRC77-061305R}.
A self-consistent and microscopic study of $Y_{32}$ effects in these $N = 150$ isotones 
was carried out in Ref.~\cite{Zhao2012_PRC86-057304}
and it was found that, for ground states of $^{248}$Cf and $^{250}$Fm,
$\beta_{32} > 0.03$ and the energy gain due to the $\beta_{32}$ distortion
is more than 300 keV.

In this work we will study the potential energy surface and tetrahedral shape
in neutron rich even-even Zr isotopes by using the covariant density function theory (CDFT).
The nuclear CDFT is one of the state-of-the-art approaches for 
the study of ground states as well as excited states in nuclei ranging from 
light to superheavy regions 
\cite{Serot1986_ANP16-1,Reinhard1989_RPP52-439,Ring1996_PPNP37-193,%
Bender2003_RMP75-121,Vretenar2005_PR409-101,Meng2006_PPNP57-470,Paar2007_RPP70-691,%
Niksic2011_PPNP66-519,Liang2015_PR570-1,Meng2015_JPG42-093101,Meng2016_WorldSci}.
Recently, the global performance of various relativistic functionals on the ground state 
properties and beyond-mean-field correlations of nuclei across the nuclear chart
has also been examined 
\cite{Afanasjev2013_PLB726-680,Agbemava2014_PRC89-054320,
Lu2015_PRC91-027304,Agbemava2015_PRC92-054310,
Afanasjev2016_PRC93-054310}. 
\label{page:global}

Recently, we have developed multidimensionally constrained CDFT (MDC-CDFT) 
by breaking the reflection and axial symmetries simultaneously.
In the MDC-CDFT, the nuclear shape is assumed to be invariant under 
the reversion of $x$ and $y$ axes; i.e., the intrinsic symmetry group 
is $V_{4}$ and all shape degrees of freedom $\beta_{\lambda\mu}$ with even $\mu$
($\beta_{20}$, $\beta_{22}$, $\beta_{30}$, $\beta_{32}$, $\beta_{40}$, $\cdots$)
are included self-consistently.
The MDC-CDFT consists of two types of models: the multidimensionally constrained
relativistic mean field (MDC-RMF) model and the multidimensionally constrained
relativistic Hartree-Bogoliubov (MDC-RHB) model.
In the MDC-RMF model, 
the BCS approach has been implemented for the particle-particle (pp) channel. 
This model has been used to study potential energy surfaces and 
fission barriers of actinides
\cite{Lu2012_PRC85-011301R,Lu2014_JPCS492-012014,Lu2014_PRC89-014323,%
Lu2014_PS89-054028,Lu2012_EPJWoC38-05003,Zhao2015_PRC91-014321},
the spontaneous fission of several fermium isotopes \cite{Zhao2016_PRC93-044315},
the $Y_{32}$ correlations in $N=150$ isotones \cite{Zhao2012_PRC86-057304}, 
and shapes of hypernuclei \cite{Lu2011_PRC84-014328,Lu2014_PRC89-044307}; 
see Refs.~\cite{Li2014_NPR_E31-253,Lu2016_RDFNS-171,Zhou2016_PS91-063008} for recent reviews.
The Bogoliubov transformation generalizes  
the BCS quasi-particle concept and provides a unified description of 
particle-hole (ph) and pp correlations on a mean-field level. 
In the MDC-RHB model, 
pairing correlations are treated by making the Bogoliubov transformation 
and
a separable pairing force of finite range 
\cite{Tian2006_CPL23-3226,Tian2009_PLB676-44,Tian2009_PRC80-024313,Niksic2008_PRC78-034318}
is adopted.
Due to its finite range nature, the dependence of the results on 
the pairing cutoff can be avoided.
The MDC-RHB model has been used to study the spontaneous fission of 
fermium isotopes \cite{Zhao2015_PRC92-064315}.

The detailed formulas for the MDC-RMF model have been presented in Ref.~\cite{Lu2014_PRC89-014323} 
and some applications of this model were reviewed in 
Refs.~\cite{Li2014_NPR_E31-253,Lu2016_RDFNS-171,Zhou2016_PS91-063008}.
In this paper, we present the formulas of the MDC-RHB model 
and its application to the study of tetrahedral shapes of 
neutron-rich Zr isotopes. 
The paper is organized as follows. 
The formalism of the MDC-RHB model is given in Sec.~\ref{sec:formalism}. 
In Sec.~\ref{sec:results} we present the numerical details and the results on 
neutron-rich even-even Zr isotopes.
A summary is given in Sec.~\ref{sec:summary}.
Some detailed formulas used in the MDC-RHB model are given in Appendices 
\ref{app:PF} and \ref{app:TM}.

\section{\label{sec:formalism}Formalism of the MDC-RHB Model}

In the CDFT, a nucleus is treated as a system of nucleons interacting
through exchanges of mesons and photons 
\cite{Serot1986_ANP16-1,Reinhard1989_RPP52-439,Ring1996_PPNP37-193,%
Bender2003_RMP75-121,Vretenar2005_PR409-101,Meng2006_PPNP57-470,Paar2007_RPP70-691,%
Niksic2011_PPNP66-519,Liang2015_PR570-1,Meng2015_JPG42-093101,Meng2016_WorldSci}.
The effects of mesons are described either by mean fields or
by point-like interactions between the nucleons
\cite{Nikolaus1992_PRC46-1757,Burvenich2002_PRC65-044308}.
The nonlinear coupling terms 
\cite{Boguta1977_NPA292-413,Brockmann1992_PRL68-3408,Sugahara1994_NPA579-557}
or the density dependence of the coupling constants 
\cite{Fuchs1995_PRC52-3043,Niksic2002_PRC66-024306}
were introduced to give correct saturation properties of nuclear matter.
Accordingly, there are four types of covariant density functionals:
the meson exchange or point-coupling nucleon interactions combined with
the nonlinear or density dependent couplings.
All these four types of functionals have been implemented in the MDC-RHB model.
In this section, we mainly present the formalism of the RHB model with 
density dependent point couplings.

The starting point of the RHB model with density dependent point couplings
is the following Lagrangian,
\begin{equation}
\begin{aligned}
\label{eq:lagrangian}
 \mathcal{L} =\ & \bar{\psi}(i\gamma_{\mu}\partial^{\mu}-M)\psi \\ 
                &-{1\over2} \alpha_S({\rho}) \rho_{S}^{2}
                 -{1\over2} \alpha_V({\rho}) j_{V}^{2}
                 -{1\over2} \alpha_{TV}({\rho}) \vec{j}_{TV}^{2} \\
                &-{1\over2} \delta_S (\partial_\nu \rho_S) (\partial^\nu \rho_S) \\
                &- e{1-\tau_3 \over 2} A_\mu j_{V}^\mu 
                 -{1\over4} F^{\mu\nu} F_{\mu\nu}, 
\end{aligned}
\end{equation}
where $M$ is the nucleon mass, $\alpha_{S}({\rho})$, $\alpha_{V}({\rho})$, 
and $\alpha_{TV}({\rho})$ are
density-dependent couplings for different channels,
$\delta_{S}$ is the coupling constant of the derivative term,
and $e$ is the electric charge.
$\rho_{S}$, $j_{V}$, and $\vec{j}_{TV}$ are the isoscalar density, 
isoscalar current, and isovector current, respectively. 

With the Green's function technique, one can derive the Dirac-Hartree-Bogoliubov 
equation \cite{Kucharek1991_ZPA339-23,Ring1996_PPNP37-193},
\begin{equation}
\label{eq:rhb}
 \int d^{3}\bm{r}^{\prime}
  \left(\begin{array}{cc} h-\lambda & \Delta \\ -\Delta^{*} & -h+\lambda \end{array}\right)
  \left(\begin{array}{c} U_{k} \\ V_{k} \end{array}\right)
 = E_{k}\left(\begin{array}{c} U_{k} \\ V_{k} \end{array}\right),
\end{equation}
where $E_{k}$ is the quasiparticle energy, $\lambda$ is the chemical
potential, and $\hat{h}$ is the single-particle Hamiltonian,
\begin{equation}
\label{eq:hamiltonian}
 \hat{h} = \bm{\alpha}\cdot[\bm{p}-\bm{V}(\bm{r})] 
         + \beta[M+S(\bm{r})]
         + V_{0}(\bm{r}) + \Sigma_R(\bm{r}),
\end{equation}
with
the scalar potential, vector potential, and rearrangement terms,  
\begin{eqnarray}
       S & = & \alpha_{S}(\rho) \rho_{S} + \delta_{S}\triangle\rho_{S}, \nonumber \\
 V^{\mu} & = & \alpha_{V}(\rho) j_{V}^{\mu} + \alpha_{TV}(\rho) j_{TV}^{\mu}\cdot\vec{\tau} 
             + e{1-\tau_3 \over 2} A^\mu, \nonumber \\
\Sigma_R & = & {1\over2} {\partial \alpha_{S} \over \partial \rho} \rho_{S}^2
	     + {1\over2} {\partial \alpha_{V} \over \partial \rho} j_{V}^2
	     + {1\over2} {\partial \alpha_{TV} \over \partial \rho} \vec{j}_{TV}^2.
\end{eqnarray}
As usual we assume that the states are invariant under the time-reversal operation, 
which means that, in the above equations, all the currents or time-odd components vanish. 
In this case the single-particle
Hamiltonian~(\ref{eq:hamiltonian}) has the time-reversal symmetry. 
For convenience two vector densities are defined as
the time-like components of the 4-currents $j_{V}$ and $j_{TV}$,
\begin{equation}
 \rho_{V} (\bm{r})=j_{v}^{0}(\bm{r}),\qquad
 \rho_{TV}(\bm{r})=j_{TV}^{0}(\bm{r}),
\end{equation}
which are the only surviving components.
The densities are related to $U_{k}$'s and $V_{k}$'s through
\begin{eqnarray}
\label{eq:densities}
 \rho_{S}(\bm{r})  & = & \sum_{k>0} V_{k}^{\dagger}(\bm{r}) \gamma_{0} V_{k}(\bm{r}), \nonumber \\
 \rho_{V}(\bm{r})  & = & \sum_{k>0} V_{k}^{\dagger}(\bm{r}) V_{k}(\bm{r}), \nonumber \\
 \rho_{TV}(\bm{r}) & = & \sum_{k>0} V_{k}^{\dagger}(\bm{r}) \tau_{3} V_{k}(\bm{r}).
\end{eqnarray}
In the above summations, the no-sea approximation is applied.

The pairing potential reads
\begin{equation}
\begin{aligned}
 \Delta_{p_{1}p_{2}}(\bm{r}_{1}\sigma_{1},\bm{r}_{2}\sigma_{2})
 = & \int d^{3}\bm{r}_{1}^{\prime} d^{3}\bm{r}_{2}^{\prime} 
      \sum_{\sigma_{1}^{\prime}\sigma_{2}^{\prime}}^{p_{1}^{\prime}p_{2}^{\prime}} \\
   & V_{p_{1}p_{2},p_{1}^{\prime}p_{2}^{\prime}}^{{\rm pp}} 
      (\bm{r}_{1}\sigma_{1}, \bm{r}_{2}\sigma_{2},
       \bm{r}_{1}^{\prime}\sigma_{1}^{\prime}, \bm{r}_{2}^{\prime}\sigma_{2}^{\prime}) \\
  & \times \kappa_{p_{1}^{\prime}p_{2}^{\prime}} 
      (\bm{r}_{1}^{\prime}\sigma_{1}^{\prime}, 
       \bm{r}_{2}^{\prime}\sigma_{2}^{\prime}),
\end{aligned}
\end{equation}
where $p=f,g$ is used to represent the large and small components of the Dirac spinor. 
$V^{{\rm pp}}$ is the effective pairing interaction
and $\kappa(\bm{r}_{1}\sigma_{1},\bm{r}_{2}\sigma_{2})$ is the pairing tensor,
\begin{equation}
 \kappa_{\alpha\alpha'}(\bm{r_1}\sigma_1,\bm{r_2}\sigma_2) 
 = \sum_{k>0} V^{*}_{\alpha k}(\bm{r_1}\sigma_1) U_{\alpha' k}(\bm{r_2}\sigma_2).
\end{equation}
As is usually done in the RHB theory, only the large components of the spinors are 
used to build the pairing potential \cite{Serra2002_PRC65-064324}. 
In practical calculations this means
\begin{equation}
\begin{aligned}
\Delta_{ff}(\bm{r}_{1}\sigma_{1},\bm{r}_{2}\sigma_{2})
= & \int d^{3} \bm{r}_{1}^{\prime}d^{3} \bm{r}_{2}^{\prime} \sum_{\sigma_{1}^{\prime}\sigma_{2}^{\prime}} \\
  & V_{ff,ff}^{{\rm pp}}(\bm{r}_{1}\sigma_{1}, \bm{r}_{2}\sigma_{2}, \bm{r}_{1}^{\prime}\sigma_{1}^{\prime}, 
    \bm{r}_{2}^{\prime}\sigma_{2}^{\prime}) \\
  & \times \kappa_{ff}(\bm{r}_{1}^{\prime}\sigma_{1}^{\prime}, \bm{r}_{2}^{\prime}\sigma_{2}^{\prime}).
\end{aligned}
\end{equation}
The other components $\Delta_{fg}$, $\Delta_{gf}$, and $\Delta_{gg}$ are omitted.

In the pp channel, we use a separable pairing force of finite range 
\cite{Tian2006_CPL23-3226,Tian2009_PLB676-44,Tian2009_PRC80-024313,Niksic2008_PRC78-034318},
\begin{eqnarray}
\label{eq:separable}
 V(\bm{r}_{1}\sigma_{1},\bm{r}_{2}\sigma_{2},\bm{r}_{1}^{\prime}\sigma_{1}^{\prime},\bm{r}_{2}^{\prime}\sigma_{2}^{\prime})
&=& -G \delta(\bm{R}-\bm{R}^{\prime}) P(r)P(r^{\prime}) \nonumber \\
 &&\times \frac{1}{2} (1-P_{\sigma}),
\end{eqnarray}
where $G$ is the pairing strength and 
$\bm{R}=(\bm{r}_{1}+\bm{r}_{2}) / 2$ and $\bm{r}=\bm{r}_{1}-\bm{r}_{2}$
are the center-of-mass and relative coordinates, respectively. 
$P(\bm{r})$ denotes the Gaussian function,
\begin{equation}
 P(\bm{r}) = {(4\pi a^2)^{-3/2}} e^{-{r^2}/{4 a^2}},
\end{equation}
where $a$ is the effective range of the pairing force.
The two parameters $G=G_0=728$ MeV$\cdot$fm$^3$ and $a=0.644$ fm 
\cite{Tian2009_PLB676-44, Tian2009_PRC80-024313}
have been adjusted to reproduce the density dependence of the 
pairing gap of symmetric nuclear matter at the Fermi surface
calculated with the Gogny force D1S.
In the present work, the pairing strength $G$ is fine-tuned to reproduce 
the pairing gaps of Zr isotopes as discussed in 
Sec.~\ref{subsec:numerical}.
\label{page:fine-tuning}

The RHB equation~(\ref{eq:rhb}) is solved by expanding the large and small components
of the spinors $U_{k}(\bm{r}\sigma)$ and $V_{k}(\bm{r}\sigma)$ in 
an axially deformed harmonic oscillator (ADHO) basis \cite{Gambhir1990_APNY198-132},
\begin{equation}
\begin{aligned}
U_{k}(\bm{r}\sigma) =& \left( \begin{array}{c}
					   \sum_{\alpha} f_{U}^{k\alpha} \Phi_{\alpha}(\bm{r}\sigma) \\ 
					   \sum_{\alpha} g_{U}^{k\alpha} \Phi_{\alpha}(\bm{r}\sigma)
				   \end{array} \right), \\
V_{k}(\bm{r}\sigma) =& \left( \begin{array}{c}
					   \sum_{\alpha} f_{V}^{k\alpha} \Phi_{\alpha}(\bm{r}\sigma) \\ 
					   \sum_{\alpha} g_{V}^{k\alpha} \Phi_{\alpha}(\bm{r}\sigma)
				   \end{array} \right),		
\end{aligned}		   			   
\end{equation}
with
\begin{eqnarray}
\label{eq:Phi}
\Phi_{\alpha}(\bm{r}\sigma) & = & C_{\alpha} \phi_{n_{z}}(z) \phi_{n_{r}}^{m_{l}}(\rho) 
    \frac{1}{\sqrt{2\pi}} e^{im_{l}\varphi} \chi_{m_{s}}(\sigma),
\end{eqnarray}
which are eigensolutions of the Schr\"odinger equation with an
ADHO potential,
\begin{equation}
\label{eq:adho}
\left[ -\frac{\hbar^{2}}{2M} \nabla^{2} + V_{B}(\rho,z) \right]
    \Phi_{\alpha}(\bm{r}\sigma) = E_{\alpha} \Phi_{\alpha}(\bm{r}\sigma),
\end{equation}
with
\begin{equation}
V_{B}(\rho,z) = \frac{1}{2} M (\omega_{\rho}^{2} \rho^{2} + \omega_{z}^{2} z^{2}).
\end{equation}
In Eq.~(\ref{eq:Phi}), 
$C_{\alpha}$ is a constant introduced for convenience, $\alpha=\{ n_{z},n_{r},m_{l},m_{s} \}$
is the collection of quantum numbers, and $\omega_{z}$ and $\omega_{\rho}$
are the oscillator frequencies along and perpendicular to the symmetry ($z$)
axis, respectively. The deformation of the basis $\beta_{{\rm B}}$
is defined through relations $\omega_{z}=\omega_{0}\exp(-\sqrt{5/4\pi}\beta_{{\rm B}})$
and $\omega_{\rho}=\omega_{0}\exp(\sqrt{5/16\pi}\beta_{{\rm B}})$,
where $\omega_{0}=(\omega_{z}\omega_{\rho}^{2})^{1/3}$ is
the frequency of the corresponding spherical oscillator. These bases
are also eigenfunctions of the $z$ component of the angular momentum
$j_{z}$ with eigenvalues $K_{\alpha}=m_{l}+m_{s}$. For such a basis
state $\Phi_{\alpha}(\bm{r}\sigma)$ we can define a time-reversal
state by $\Phi_{\bar{\alpha}}(\bm{r}\sigma)=\mathcal{T} \Phi_{\alpha}(\bm{r}\sigma)$,
where $\mathcal{T}=i\sigma_{y}K$ is the time-reversal operator and
$K$ is the complex conjugation. It is easy to see that $\Phi_{\bar{\alpha}}(\bm{r}\sigma)$
is also an eigensolution of Eq.~(\ref{eq:adho}) with the same energy $E_{\alpha}$,
while the direction of the angular momentum is reversed, $K_{\bar{\alpha}}=-K_{\alpha}$.
$\{\Phi_{\alpha},\Phi_{\bar{\alpha}}|K_{\alpha}>0\}$
forms a complete and discrete basis set in the space of two-component spinors.

In practical calculations, the ADHO basis is truncated as
\cite{Warda2002_PRC66-014310,Lu2014_PRC89-014323},
\begin{equation}
\label{eq:truncation}
 \left[ n_{z} / Q_{z} + (2n_{\rho}+|m|) / Q_{\rho} \right] \leq N_{f},
\end{equation}
for the large component of the Dirac spinor. 
$N_{f}$ is a certain integer constant and $Q_{z}=\max(1,b_{z}/b_{0})$
and $Q_{\rho}=\max(1,b_{\rho}/b_{0})$ are constants calculated from
the oscillator lengths $b_{0}=1/\sqrt{M\omega_0}$, $b_{z}=1/\sqrt{M\omega_z}$, 
and $b_{\rho}=1/\sqrt{M\omega_\rho}$.
For the small component, the truncation is made up to $N_g=N_f+1$ major 
shells in order to avoid spurious states \cite{Gambhir1990_APNY198-132}.
The convergence of the results on $N_{f}$ is discussed in Sec.~\ref{subsec:numerical}.

The $V_4$ symmetry is imposed in the MDC-CDFT \cite{Zhou2016_PS91-063008}; 
i.e., the nuclear potentials and densities are invariant under the following operations
\begin{equation}
\begin{aligned}
\label{eq:v4}
\hat{S}_x \phi(x,y,z)  =& \ \phi(-x,y,z), \\
\hat{S}_y \phi(x,y,z) =& \ \phi(x,-y,z), \\
\hat{S} \phi(x,y,z) =& \ \phi(-x,-y,z),
\end{aligned}
\end{equation}
where $\phi(x,y,z)$ represents nuclear potentials and densities. 
Thus, both axial symmetry and reflection symmetry are broken.
Due to the $V_4$ symmetry, 
one can introduce a simplex operator $\hat{S}=ie^{-i\pi j_{z}}$ 
to block-diagonalize the RHB matrix.
For a fermionic system, $\hat{S}$ is a Hermitian operator 
and satisfies $\hat{S}^{2}=1$. 
The basis states are also eigenstates of $\hat{S}$ with $\hat{S} 
\Phi_{\alpha} = (-1)^{K_{\alpha}-\frac{1}{2}}\Phi_{\alpha}$,
which means that $\Phi_{\alpha}$ with $K_{\alpha} = 
\frac{1}{2}, -\frac{3}{2}, \frac{5}{2}, -\frac{7}{2}, \cdots$
span a subspace with $S=+1$, while their time-reversal states span the one with $S=-1$.
Due to the time-reversal symmetry, the RHB matrix is block-diagonalized into 
two smaller ones denoted by the quantum number $S=\pm 1$, respectively. 
Furthermore, for a system with the time-reversal symmetry, 
it is only necessary to diagonalize the matrix with $S=+1$ and 
the other half is obtained by making the time reversal operation on Dirac spinors.

In the $S=+1$ subspace, solving the RHB equation~(\ref{eq:rhb}) is equivalent
to diagonalizing the matrix
\begin{eqnarray}
\left( \begin{array}{cc} \mathcal{A}-\lambda & \mathcal{B} \\ \mathcal{B}^{\dagger} 
& -\mathcal{A}^{*}+\lambda \end{array} \right)
\left( \begin{array}{c} \mathcal{U}_{k} \\ \mathcal{V}_{k} \end{array} \right) 
& = & E_{k} \left( \begin{array}{c} \mathcal{U}_{k} \\ \mathcal{V}_{k} \end{array} \right),
\end{eqnarray}
where $\mathcal{U}_{k}=(\mathcal{U}_{k,\alpha})$ and
      $\mathcal{V}_{k}=(\mathcal{V}_{k,\bar{\alpha}})$
are column matrices, and
\begin{eqnarray}
\mathcal{A}_{\alpha\alpha^{\prime}} = \langle \alpha|h|\alpha^{\prime} \rangle,\qquad
\mathcal{B}_{\alpha\bar{\alpha^{\prime}}} = \langle \alpha|\Delta|\bar{\alpha^{\prime}} \rangle, 
\end{eqnarray}
are matrix elements of the single-particle Hamiltonian and the pairing
field, where $\alpha$ and $\alpha^{\prime}$ run over all the quantum
numbers $\left\{ n_{z},n_{r},m_{l},m_{s},p\right\} $ satisfying the
truncation condition (\ref{eq:truncation}).

We expand the potentials $V(\bm{r})$ and $S(\bm{r})$ and 
the densities in Eq.~(\ref{eq:densities}) in terms of the Fourier series,
\begin{equation}
\label{eq:fourier0}
 f(\rho,\varphi,z) = \sum_{\mu=-\infty}^{\infty} f_{\mu}(\rho,z) 
                      \frac{1}{\sqrt{2\pi}} \exp(i\mu\varphi).
\end{equation}
By applying the symmetry conditions (\ref{eq:v4}),
it is easy to see that $f_{\mu}=f_{\mu}^{*}=f_{\bar{\mu}}$ and $V_{n}=0$ for odd $n$. 
Thus the expansion in Eq.~(\ref{eq:fourier0}) can be simplified as
\begin{equation}
\label{eq:fourier}
f(\rho,\varphi,z) = f_{0}(\rho,z) \frac{1}{\sqrt{2\pi}} + \sum_{n=1}^{\infty} f_{n}(\rho,z) \frac{1}{\sqrt{\pi}} \cos(2n\varphi),
\end{equation}
where
\begin{eqnarray}
\label{eq:fcof}
f_{0}(\rho_{,}z) & = & \frac{1}{\sqrt{2\pi}} \int_{0}^{2\pi}d\varphi f(\rho,\varphi,z), \nonumber \\
f_{n}(\rho,z) & = & \frac{1}{\sqrt{\pi}} \int_{0}^{2\pi}d\varphi f(\rho,\varphi,z) \cos(2n\varphi),
\end{eqnarray}
are real functions of $\rho$ and $z$. 
The calculation of matrix elements $\mathcal{A}_{\alpha\alpha^{\prime}}$ and densities
are similar to that of the MDC-RMF model; 
the details can be found in Appendices of Ref.~\cite{Lu2014_PRC89-014323}.

Due to the finite range nature of the paring force given in Eq.~(\ref{eq:separable}), 
it is not easy to calculate the matrix element $V_{12,1^{\prime}2^{\prime}}^{{\rm pp}}$,
where the numbers $1,2,1^{\prime}$, and $2^{\prime}$ denote the ADHO
basis wave functions. 
In Ref.~\cite{Tian2009_PRC80-024313}, the authors have shown that the anti-symmetrized
matrix element $\bar{V}_{12,1^{\prime}2^{\prime}}^{{\rm pp}} 
= V_{12,1^{\prime}2^{\prime}}^{{\rm pp}} - V_{12,2^{\prime}1^{\prime}}^{{\rm pp}}$ 
can be written as a sum of separable terms in both axially deformed
and anisotropic HO bases. For this purpose the first step is to represent
the product of two single-particle HO wave functions as a sum of
HO wave functions in their center-of-mass frame,
\begin{eqnarray}
\label{eq:ppho}
|n_{r_{1}}m_{1}\rangle |n_{r_{2}}m_{2}\rangle & = & \sum_{N_{p}M_{p}} \sum_{n_{p}m_{p}}
    M_{N_{p}M_{p}n_{p}m_{p}}^{n_{r_{1}}m_{1}n_{r_{2}}m_{2}} |N_{p}M_{p}\rangle |n_{p}m_{p}\rangle, \nonumber \\
|n_{z_{1}}\rangle |n_{z_{2}}\rangle & = & \sum_{N_{z}n_{z}}M_{N_{z}n_{z}}^{n_{z_{1}}n_{z_{2}}} 
    |N_{z}\rangle |n_{z}\rangle,
\end{eqnarray}
where $M_{N_{z}n_{z}}^{n_{z_{1}}n_{z_{2}}}$ and $M_{N_{p}M_{p}n_{p}m_{p}}^{n_{r_{1}}m_{1}n_{r_{2}}m_{2}}$
are the Talmi-Moshinski brackets which are given in Appendix~\ref{app:TM}.
The matrix element in the center-of-mass frame reads
\begin{eqnarray}
\label{eq:pmatrix}
V_{12,1^{\prime}2^{\prime}} & = & -2\sqrt{2} G \sum_{N_{z}N_{p}M_{p}} \left(W_{12}^{N_{z}N_{p}M_{p}}\right)^* W_{1^{\prime}2^{\prime}}^{N_{z}N_{p}M_{p}},
\end{eqnarray}
where
\begin{eqnarray}
W_{12}^{N_{z}N_{p}M_{p}} &= & \delta_{K_{1}+K_{2},M_{p}} \delta_{\pi_{1}\pi_{2},(-1)^{N_{z}+|M_{p}|}} 
    \tau_{1} \frac{1}{\sqrt{2}} C_{1} C_{2}  \nonumber \\
 &&\times
    \left( \sum_{n_{z}} M_{N_{z}n_{z}}^{n_{z_{1}}n_{z_{2}}} V_{n_{z}} \right) \nonumber \\
 &&\times
    \left( \sum_{n_{p}} M_{N_{p}M_{p}n_{p}0}^{n_{r_{1}}m_{1}n_{r_{2}}m_{2}} U_{n_{p}} \right),
\end{eqnarray}
and
\begin{eqnarray}
V_{n_{z}} & = & \frac{1}{(4\pi a^{2})^{1/2}} \int_{-\infty}^{\infty}dz e^{-\frac{z^{2}}{2a^{2}}} \phi_{n_{z}}\left(z\right), \nonumber \\
U_{n_{p}} & = & \frac{\sqrt{2\pi}}{4\pi a^{2}} \int_{0}^{\infty}d\rho \rho e^{-\frac{\rho^{2}}{2a^{2}}} R_{n_{p}}^{0} \left(\rho\right) .
\end{eqnarray}
More details are given in Appendix~\ref{app:PF}.
The pairing field and pairing energy can also be written in the separable form as
\begin{eqnarray}
\label{eq:delta}
\Delta_{12} & = & \sum_{1^{\prime}2^{\prime}} V_{12,1^{\prime}2^{\prime}} \kappa_{1^{\prime}2^{\prime}} \nonumber \\
            & = & -2\sqrt{2}G\sum_{N_{z}} \sum_{N_{p}M_{p}} \left(W_{12}^{N_{z}N_{p}M_{p}}\right)^* 
                                                                  P^{N_{z}N_{p}M_{p}},  \\
E_{{\rm pair}} & = & \frac{1}{2} \sum_{12,1^{\prime}2^{\prime}} V_{12,1^{\prime}2^{\prime}} 
                \kappa_{12}^{*}\kappa_{1^{\prime}2^{\prime}} \nonumber \\
               & = & -\sqrt{2} G \sum_{N_{z}} \sum_{N_{p}M_{p}} |P^{N_{z}N_{p}M_{p}}|^{2},
\end{eqnarray}
where
\begin{equation}
P^{N_{z}N_{p}M_{p}} = \sum_{12} W_{12}^{N_{z}N_{p}M_{p}} \kappa_{12}.
\end{equation}

Since $\hat{S}=ie^{-i\pi j_{z}}$ is conserved, 
we can check that the matrix element of the pairing tensor has the structure 
$\kappa_{12}=\delta_{S_{1},\bar{S_{2}}} \kappa_{12}$.
Combined with the $\delta_{K_{1}+K_{2},M_{p}}$ in $W_{12}^{N_{z}N_{p}M_{p}}$,
one gets
\begin{eqnarray}
P^{N_{z}N_{p}M_{p}} & = & \sum_{12} \delta_{S_{1},\bar{S_{2}}} \delta_{K_{1}+K_{2},M_{p}} W_{12}^{N_{z}N_{p}M_{p}} \kappa_{12} \nonumber \\
                    & = & \delta_{M_{p},2n}P^{N_{z}N_{p}M_{p}},\quad n\in Z,
\end{eqnarray}
i.e., only $P^{N_{z}N_{p}M_{p}}$ with even $M_{p}$ 
survive and thus we can skip safely those with odd $M_{p}$ in the sum in Eq.~(\ref{eq:delta}). 
The sum in Eq.~(\ref{eq:pmatrix}) runs over the ADHO quantum numbers $N_{z}$, $N_{p}$, and $M_{p}$ 
in the center-of-mass frame with $N_{z}+2N_{p}+|M_{p}| \leq 2N_{f}$.

The total energy of the nucleus reads
\begin{eqnarray}
E_{{\rm total}} &= & \int d^{3}\bm{r} \left\{ \sum_{k} v_{k}^{2} \psi_{k}^{\dagger} 
      \left( \bm{\alpha} \cdot \bm{p} + \beta M_{{\rm B}} \right) \psi_{k} \right.  \nonumber \\
 & & +\ \frac{1}{2} \alpha_{S}\rho_{S}^{2} + \frac{1}{2}\alpha_{V}\rho_{V}^{2} 
    + \frac{1}{2}\alpha_{TS}\rho_{TS}^{2} + \frac{1}{2}\alpha_{TV}\rho_{TV}^{2} \nonumber \\
 & & +\ \frac{1}{3}\beta_{S}\rho_{S}^{3} + \frac{1}{4}\gamma_{S}\rho_{S}^{4} + \frac{1}{4}\gamma_{V}\rho_{V}^{4} \nonumber \\
  && +\ \frac{1}{2}\delta_{S}\rho_{S}\Delta\rho_{S} + \frac{1}{2}\delta_{V}\rho_{V}\Delta\rho_{V} \nonumber \\
 & &\left.  +\ \frac{1}{2}\delta_{TS}\rho_{TS}\Delta\rho_{TS} 
    + \frac{1}{2}\delta_{TV}\rho_{TV}\Delta\rho_{TV} + \frac{1}{2}e\rho_{C}A \right\} \nonumber \\
 & & +\ E_{{\rm pair}} + E_{{\rm c.m.}}, 
\end{eqnarray}
where the center-of-mass correction $E_{{\rm c.m.}}$ is calculated,
depending on the relativistic functional,
either phenomenologically as
\begin{equation}
 E_{{\rm c.m.}} = - \frac{3}{4} \times 41 A^{1/3} \ \textrm{MeV},
\end{equation}
or from the quasiparticle vacuum
\begin{equation}
 E_{{\rm c.m.}} = -\frac{\langle P^{2}\rangle}{2MA},
\end{equation}
where $P$ is the total linear momentum and $A$ is the nuclear mass
number.

The intrinsic multipole moments are calculated as
\begin{equation}
Q_{\lambda\mu}^{\tau} = \int d^{3}\bm{r}\rho_{V}^{\tau}(\bm{r})r^{\lambda}Y_{\lambda\mu}(\Omega),
\end{equation}
where $Y_{\lambda\mu}(\Omega)$ is the spherical harmonics and $\tau$
refers to the proton, neutron, or the whole nucleus. The deformation
parameter $\beta_{\lambda\mu}$ is obtained from the corresponding
multipole moment by
\begin{equation}
\beta_{\lambda\mu}^{\tau}=\frac{4\pi}{3N_{\tau}R^{\lambda}}Q_{\lambda\mu}^{\tau},
\end{equation}
where $R=1.2 \times A^{1/3}$ fm
and $N_{\tau}$ is the number of proton, neutron, or nucleons.

\section{Results and discussions}\label{sec:results}

In this section, we study neutron rich even-even Zr isotopes with $100 \leq A \leq 114$ 
by using the MDC-RHB model.
These nuclei are among those with $N \simeq 70$ and $A \simeq 110$ which show
interesting nuclear structure properties, such as drastic changes in
shapes with $A$ and shape coexistences 
\cite{Skalski1997_NPA617-282,Lalazissis1999_ADNDT71-1,Xu2002_PRC65-021303R,
Dudek2002_PRL88-252502,
Geng2003_PTP110-921,
Schunck2004_PRC69-061305R,
Schunck2004_PRC69-061305R,Olbratowski2006_IJMPE15-333,
Verma2008_PRC77-024308,
Zberecki2009_PRC79-014319,
Bender2009_PRC80-064302,
Rodriguez-Guzman2010_PLB691-202,
Dong2010_ScichinaPMA53-106,Dong2011_CTP56-922,
Mali2011_IJMPE20-2293,
Liu2011_NPA858-11,
Shi2012_PRC85-027307,
Xiang2012_NPA873-1,Mei2012_PRC85-034321,
Liu2013_JPCS420-012046,
Agbemava2014_PRC89-054320,
Liu2015_SciChinaPMA58-112003,
Togashi2016_PRL117-172502,
Xiang2016_PRC93-054324,
Kremer2016_PRL117-172503,Nomura2016_PRC94-044314}.
The structures of nuclei around $N=70$ are also of particular interest for 
nuclear astrophysics because neutron-rich nuclei with $A \simeq 110$
are around the $r$-process path, which is determined 
by the equilibrium between neutron capture and photodisintegration
\cite{Kratz1993_ApJ403-216,Dobaczewski1994_PRL72-981,
Chen1995_PLB355-37,Pearson1996_PLB387-455,%
Dobaczewski1996_PRC53-2809,Pfeiffer1996_APPB27-475,%
Pfeiffer2001_NPA693-282,Wanajo2006_NPA777-676,Qian2007_PR442-237,Kratz2007_PPNP59-147,
Nishimura2011_PRL106-052502}.

Zr isotopes have been studied extensively with the CDFT under the 
assumption of reflection symmetry 
\cite{Lalazissis1999_ADNDT71-1,Geng2003_PTP110-921,Mali2011_IJMPE20-2293,
Xiang2012_NPA873-1,Mei2012_PRC85-034321,
Agbemava2014_PRC89-054320,
Xiang2016_PRC93-054324}.
In these studies, the coexistence of prolate and oblate minima in 
potential energy curves of even-even Zr isotopes were discussed.
By including the $\beta_{30}$ deformation, a pear-like ground state of 
$^{112}$Zr is predicted with the functional DD-PC1 \cite{Agbemava2016_PRC93-044304}.
From the experimental point of view,
the analysis of rotational bands, $\beta$-decay half-lives, 
and lifetimes of the first $2^+$ states reveals that some of these nuclei 
are well deformed
\cite{Nishimura2011_PRL106-052502,Hwang2006_PRC74-017303,Mach1990_PRC41-350,%
Pereira2009_PRC79-035806,Hwang2006_PRC73-044316,Cheifetz1970_PRL25-38,%
Sumikama2011_PRL106-202501,Hua2004_PRC69-014317,Sarriguren2010_PRC81-064314,%
Sarriguren2014_PRC89-034311,Browne2015_PLB750-448,Ni2014_PRC89-064320,%
Yeoh2010_PRC82-027302,Li2008_PRC78-044317}
and there exist shape coexistences in $^{100-104}$Zr 
\cite{Hwang2006_PRC74-017303,Mach1990_PRC41-350,Pereira2009_PRC79-035806,%
Petrovici2013_JPCS413-012007}.

Next we give some numerical details and illustrative calculations of 
the MDC-RHB model. 
Then we present and discuss the results for neutron-rich even-even Zr isotopes
calculated with functionals DD-PC1 \cite{Niksic2008_PRC78-034318} 
and PC-PK1 \cite{Zhao2010_PRC82-054319}.

\subsection{\label{subsec:numerical}Numerical details}

\begin{table}
\centering
\caption{\label{tab:gaps} %
Pairing gaps (in MeV) calculated from functionals DD-PC1 \cite{Niksic2008_PRC78-034318} 
and PC-PK1 \cite{Zhao2010_PRC82-054319} for $^{102,104}$Zr.
The experimental values are obtained from odd-even mass differences
and the mass values are taken from Ref.~\cite{Wang2012_ChinPhysC36-1603}.
}
\begin{ruledtabular}
\begin{tabular}{lcccc}
            & \multicolumn{2}{c}{$^{102}$Zr} & \multicolumn{2}{c}{$^{104}$Zr} \\ \hline
            & $\Delta_n$     & $\Delta_p$    & $\Delta_n$    & $\Delta_p$     \\ \hline
 Experiment & 1.10           & 1.55          & 1.08          & 1.55           \\ \hline
 DD-PC1     & 1.08           & 1.51          & 0.99          & 1.50           \\ \hline
 PC-PK1     & 1.17           & 1.60          & 1.11          & 1.60           \\ 
\end{tabular}
\end{ruledtabular}
\end{table}

In the MDC-RHB model, the potentials and densities are calculated in a spatial lattice 
in which mesh points in the $\rho$ and $z$ directions are designed in a way that 
the Gaussian quadrature can be performed and those for the azimuthal angle $\varphi$
are equally distributed. 
Since we keep the mirror reflection symmetry with respect to the $x = 0$ or $y = 0$ planes, 
only mesh points with positive $x$ and $y$ are considered.
For axially deformed nuclei the azimuthal degree of freedom vanishes, 
and for reflection symmetric nuclei mesh points with $z < 0$ can also be omitted.  
The values of the localized fields and potentials in the full lattice space 
can be simply obtained 
by symmetry transformations such as rotations or the spatial reflection. 

To reproduce the available empirical pairing gaps of $^{102,104}$Zr 
(see Table~\ref{tab:gaps}),
the strength of the pairing force given in Eq.~(\ref{eq:separable})
is readjusted a little for protons compared to those originally
proposed in Refs.~\cite{Tian2009_PLB676-44, Tian2009_PRC80-024313}: 
$G_n=G_0$, $G_p=1.12G_0$ with $G_0=738$ MeV$\cdot$fm$^3$.

\begin{figure}
\includegraphics[width=0.48\textwidth]{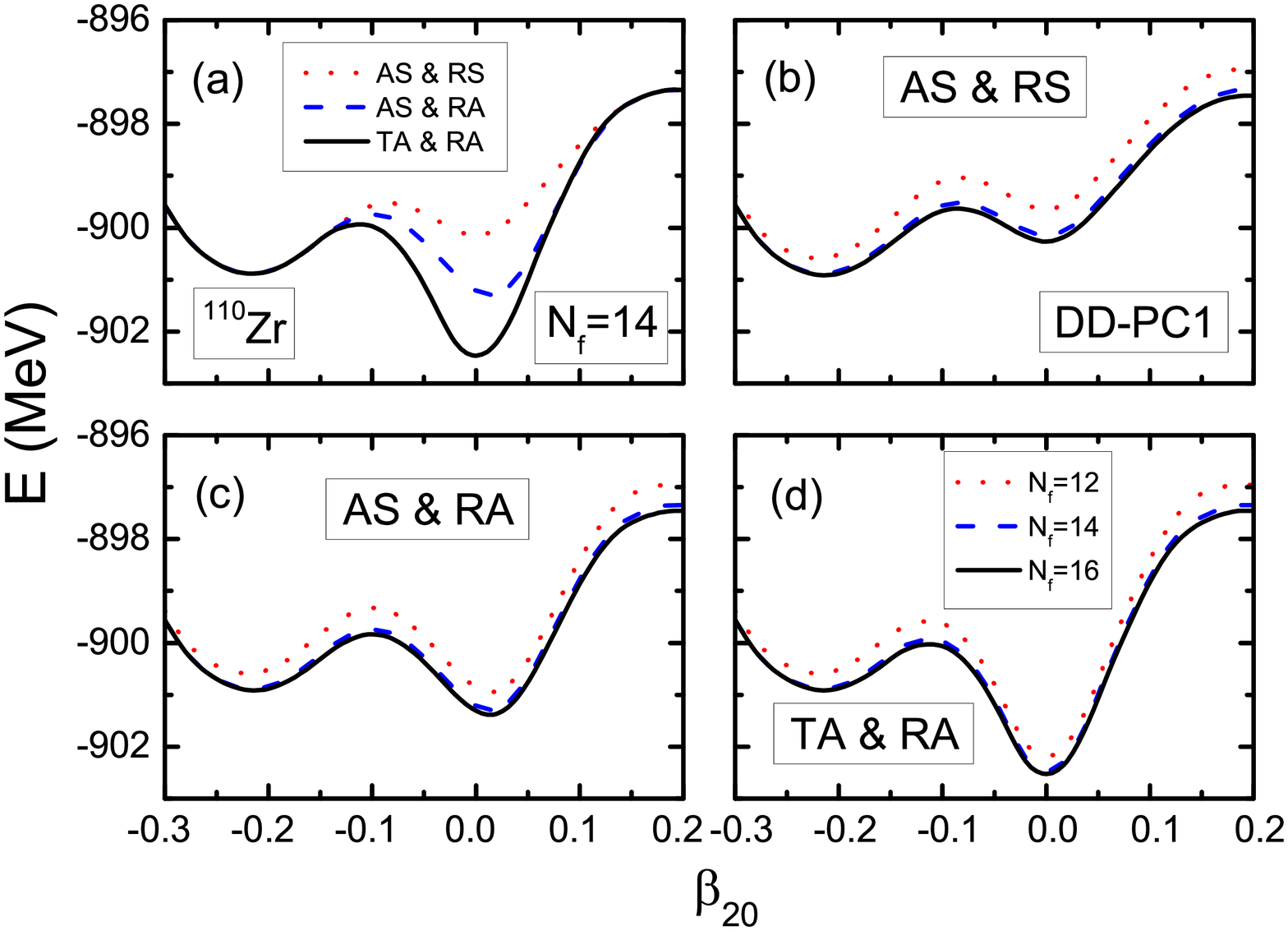}
\includegraphics[width=0.44\textwidth]{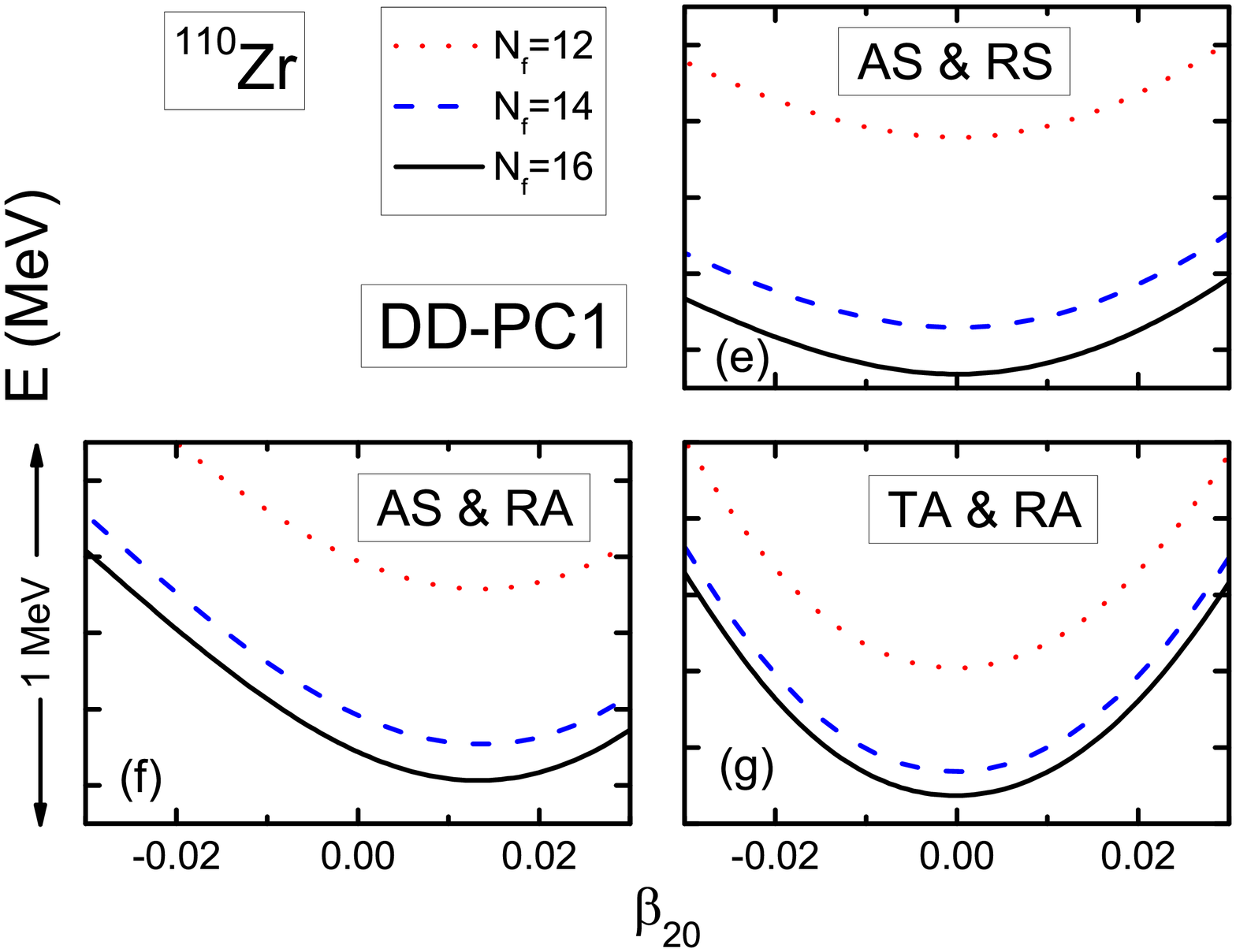}
\caption{(Color online)~\label{fig:Zr110_Nf} %
The potential energy curve of $^{110}$Zr from MDC-RHB calculations 
with different symmetries imposed (a) and
with different truncations of the ADHO basis [(b), (c), and (d)]. 
The calculations are preformed with axial symmetry
and reflection symmetry (AS \& RS), axial symmetry and reflection asymmetry (AS \& RA), 
and triaxial symmetry and reflection asymmetry (TA \& RA) imposed, respectively. 
In [(b), (c), and (d)], the results calculated with $N_f = 12$, 14, and 16 are depicted by dotted, 
dashed, and solid curves, respectively. 
The corresponding amplified figures [(e), (f), and (g)] show the detailed structure of the PES's
near $\beta_{20}=0$; the width of each subfigure is 0.06 and the height is 1 MeV.
} 
\end{figure}

The calculated physical observables should converge as the truncation $N_f \rightarrow \infty$. 
In Fig.~\ref{fig:Zr110_Nf}, we show the one-dimensional (1D) potential energy curve 
of $^{110}$Zr around the ground state with 
axial symmetry and reflection symmetry, 
axial symmetry and reflection asymmetry, and 
nonaxial and reflection asymmetry imposed, respectively.
The effective interaction DD-PC1 \cite{Niksic2008_PRC78-034318} is used.
Calculations with different truncations ($N_f=12$, 14, and 16) are depicted by 
dotted, dashed, and solid curves, respectively.
As shown in the upper panel of Fig.~\ref{fig:Zr110_Nf}, 
the predicted ground state shape of $^{110}$Zr is oblate with $\beta_{20} \approx -0.2$ 
when both axial and reflection symmetries are imposed.
There is another minimum around $\beta_{20} = 0$.
The energy of this minimum is reduced by about 1 MeV when the axial octupole deformation 
($\beta_{30}$) is included,
while the energy gain is about 2 MeV when the nonaxial octupole deformation 
($\beta_{32}$) is included.
As a result, the ground state changes to be $\beta_{20} \approx 0$.
To see more clearly the truncation errors, 
we have amplified the figure near $\beta_{20} = 0$.
We can see that for $^{110}$Zr, the energy changes about 400 keV 
when $N_f$ increases from 12 to 14; further increasing $N_f$ from 14 to 16, 
the energy changes only about 100 keV. 
Furthermore, from Fig.~\ref{fig:Zr110_Nf}, 
one can find that the inaccuracy caused by the finite 
basis size is almost independent of the imposed symmetry 
(in this case, independent of whether the axial or nonaxial deformations, 
reflection symmetric or reflection asymmetric deformations are included).
This means a good convergence and $N_f=14$ is sufficient in this mass region. 

\begin{figure}
\includegraphics[width=0.45\textwidth]{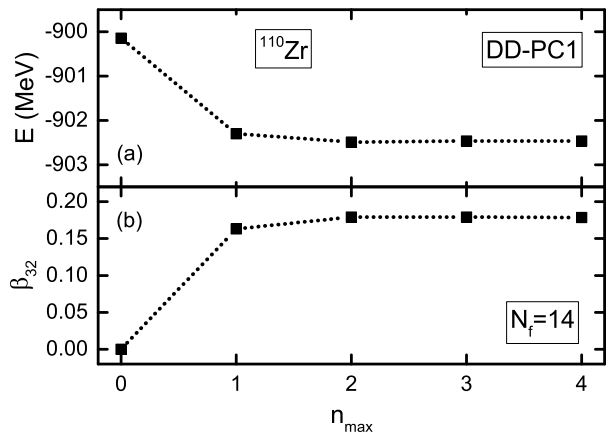}
\caption{\label{fig:Zr110_Ncov} %
The binding energy (a) and $\beta_{32}$ deformation (b) of $^{110}$Zr 
from MDC-RHB calculations with different truncations made in the Fourier expansion 
of densities and potentials. 
The calculations are performed with $N_f = 14$. 
The functional DD-PC1 is used in the MDC-RHB calculations.
} 
\end{figure}

For axially symmetric nuclei where potentials and densities 
$f(\rho,\varphi,z)\equiv f(\rho,z)$ are independent of the azimuthal angle $\varphi$, 
the Fourier coefficient $f_n(\rho,z) = 0$ when $n \geq 1$. 
For nonaxial nuclei, $f_n(\rho,z) \neq 0$ when $n \geq 1$.
Therefore, $n$ characterizes the nonaxiality of potentials and densities.
In practical calculations, the Fourier series in Eq.~(\ref{eq:fourier}) has to be 
truncated, i.e., $n \leq n_{\rm{max}}$.
To study the dependence of the calculated results on $n_{\rm{max}}$, 
we show the binding energy and nonaxial $\beta_{32}$ deformation parameter of 
$^{110}$Zr calculated with different $n_{\rm{max}}$ in Fig.~\ref{fig:Zr110_Ncov}.
As shown in Fig.~\ref{fig:Zr110_Ncov}(b), if all $f_n$'s with $n \geq1$ 
are omitted ($n_{\rm{max}}=0$), the nucleus remains axial and $\beta_{32}=0$.
The convergence of the binding energy and $\beta_{32}$ with $n_{\rm{max}}$ increasing 
is very fast. 
The results obtained with $n_{\rm{max}}=2$ are very precise already; 
the energy gain when $n_{\rm{max}}$ increases from 2 to 3 is only around 30 keV.   
In practical calculations, $n_{\rm{max}}$ is determined automatically 
by the relation $2n_{\rm{max}} = K_{\rm{max}} - K_{\rm{min}}$, 
where $K_{\rm{max}}$  ($K_{\rm{min}}$) is the largest (smallest) $K$ 
in the truncated ADHO basis.
This is sufficiently large for all of results discussed here.

\subsection{\label{subsec:zr}Neutron-rich even-even Zr isotopes}

\begin{figure*}
 \includegraphics[width=0.85\textwidth]{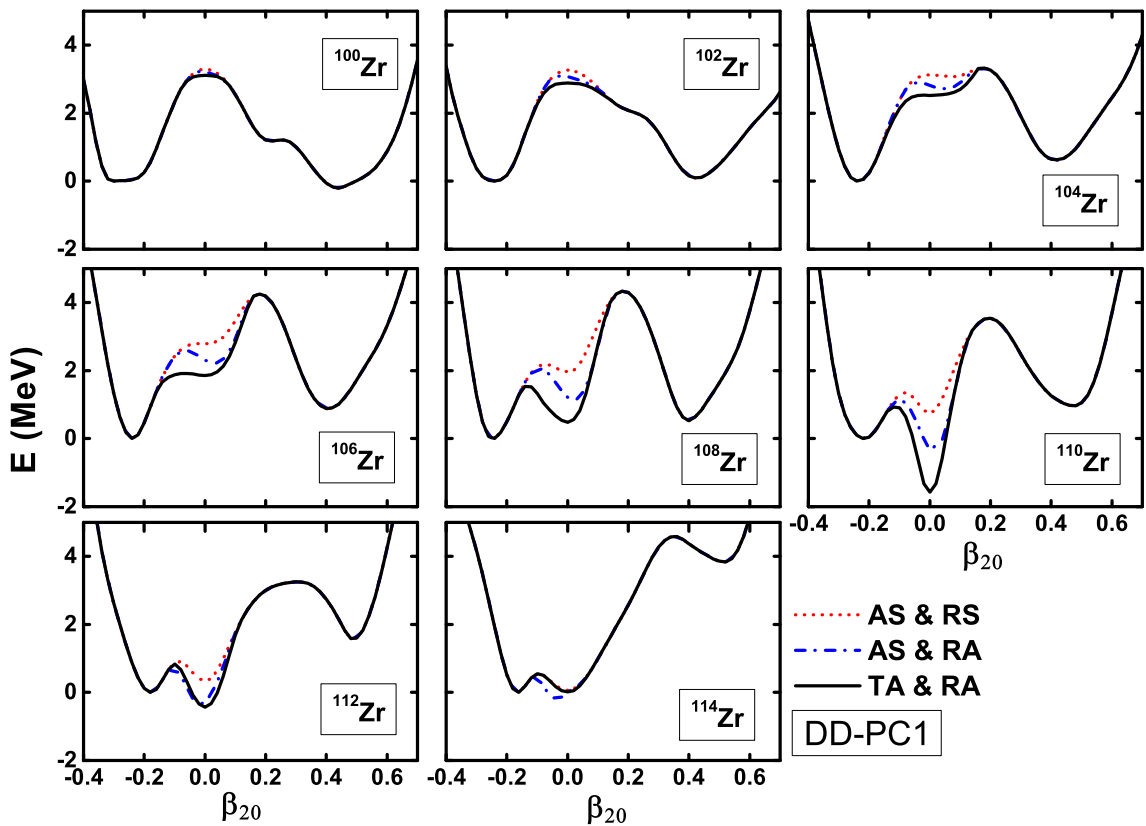}
\caption{(Color online)~\label{fig:Zr_DDPC1}%
Potential energy curves ($E\sim \beta_{20}$) for Zr isotopes.
The energy is normalized with respect to the energy minimum with $\beta_{20}<0$ for each nucleus.
The dotted line denotes results with axial symmetry (AS) and reflection symmetry (RS) imposed,
the dash-dotted line denotes results restricted to AS and reflection asymmetry (RA),
while the solid line represents results with both AS and RS broken, i.e.,
with triaxial (TA) deformation and RA.
The functional DD-PC1 is used in MDC-RHB calculations.
}
\end{figure*}

We calculate one-dimensional (1D) potential energy curves ($E\sim\beta_{20}$) 
for even-even Zr nuclei with $100 \leq A \leq 114$.
The functionals DD-PC1 \cite{Niksic2008_PRC78-034318} and 
PC-PK1 \cite{Zhao2010_PRC82-054319} are used. 
To investigate different roles played by the nonaxiality and reflection asymmetry,
calculations are preformed with different symmetries imposed: 
(i) axial and reflection symmetry, 
(ii) axial symmetry and reflection asymmetry, and 
(iii) nonaxial and reflection asymmetry,
and the results are
denoted by dotted, dash-dotted, and solid lines, respectively.

First we focus on 1D potential energy curves obtained with DD-PC1, 
which are shown in Fig.~\ref{fig:Zr_DDPC1}.
If nuclear shapes are restricted to be axial and reflection symmetric
(red dotted line), 
a prolate ground state with $\beta_{20} \approx 0.4$ coexisting 
with an oblate minimum with $\beta_{20} \approx -0.2$ is obtained 
for $^{100}$Zr.
This is consistent with the experimental observations 
\cite{Mach1990_PRC41-350,Hwang2006_PRC74-017303,Hua2004_PRC69-014317,Cheifetz1970_PRL25-38}.
For $^{102-114}$Zr, the ground state shapes are predicted to be oblate with  
$\beta_{20} \approx -0.2$.
As neutron number increases, the energy of the minimum with $\beta_{20}>0$ increases.
A minimum with $\beta_{20} \approx 0$ is developed starting from $^{106}$Zr. 
This minimum becomes lower in energy and the pocket becomes deeper with
the neutron number increasing.

If nuclei are allowed to be reflection asymmetric but axial symmetric, 
one obtains 1D potential energy curves, denoted by blue dash-dotted lines
in Fig.~\ref{fig:Zr_DDPC1}.
The $\beta_{30}$ deformation is involved around $\beta_{20} =0$ for 
all of the nuclei investigated here.
As a result, a minimum develops for $^{104}$Zr around $\beta_{20} =0$.
The energy of the minimum with $\beta_{20} \approx 0$ for $^{106-114}$Zr 
is lowered substantially by the $\beta_{30}$ distortion.
Due to this lowering effect, the energy of the minimum with pear-like 
shape ($\beta_{20} \approx 0$, $\beta_{30} \neq 0$) is lower than 
the minimum with oblate or prolate shape for $^{110}$Zr, $^{112}$Zr, and $^{114}$Zr.
This lowering effect is also found in Skyrme HFB+BCS calculations \cite{Zberecki2009_PRC79-014319}.
The $\beta_{30}$ deformation is not important either around the oblate minimum 
($\beta_{20} \approx -0.2$) or around the prolate minimum ($\beta_{20} \approx 0.4$).
Note that these results are different from those presented in 
Ref.~\cite{Agbemava2016_PRC93-044304} where the octupole deformation is found 
only in the ground state of $^{112}$Zr with the functional DD-PC1. 
Such differences are mainly caused by different pairing strengths used 
in these two works.
We have checked that if we use the pairing strengths given
in Ref.~\cite{Agbemava2016_PRC93-044304}, i.e., 
$G_n=G_p=1.12G_0$ with $G_0=738$ MeV$\cdot$fm$^3$,
we can get the same results as 
Ref.~\cite{Agbemava2016_PRC93-044304}.

When the $\beta_{32}$ deformation is allowed in the calculations, 
both axial and reflection symmetries are broken.
As seen in Fig.~\ref{fig:Zr_DDPC1}, the $\beta_{32}$ deformation changes very much 
the energy around $\beta_{20}=0$ (the black solid line).
Similar to the cases discussed in the previous paragraph about the inclusion of $\beta_{30}$, 
a minimum also develops for $^{104}$Zr around $\beta_{20} =0$.
However, the $\beta_{32}$ distortion effect is more pronounced than 
that of the $\beta_{30}$ deformation for most of these nuclei.
The energy of the minimum with $\beta_{20} \approx 0$ for $^{106-112}$Zr is lowered substantially.
A tetrahedral ground state is predicted for $^{110,112}$Zr. 
For $^{114}$Zr, the predicted pear-like shape is lower in energy 
than the tetrahedral shape.
From Fig.~\ref{fig:Zr_DDPC1}, we know that the $\beta_{32}$ distortion effect 
is the most pronounced for $^{110}$Zr, 
where the inclusion of the $\beta_{32}$ deformation lowers the energy of 
the minimum around $\beta_{20} = 0$ by about 2 MeV.
Similar to the situation of $\beta_{30}$, the $\beta_{32}$ deformation is 
important neither 
around the oblate minimum ($\beta_{20} \approx -0.2$) nor 
around the prolate minimum ($\beta_{20} \approx 0.4$).
The obtained ground state deformation parameters as well as the binding energies 
with DD-PC1 for $^{110-114}$Zr are listed in Table~\ref{tab:DDPC1_Gro}.
The corresponding available experimental values are also included for comparison.   
The predicted energies of various minima 
(with respect to the corresponding ground state energy) are presented in 
Table~\ref{tab:mini_DDPC1}.  

\begin{table}
\caption{\label{tab:DDPC1_Gro} %
The quadrupole deformation $\beta_{20}$, axial octupole deformation $\beta_{30}$,
nonaxial octupole deformation $\beta_{32}$, and hexadecapole deformation $\beta_{40}$ 
together with binding energies $E_\mathrm{cal}$ for the ground states of
Zr isotopes obtained in MDC-RHB calculations with the functional DD-PC1.
$E_\mathrm{exp}$ denotes experimental binding energies taken
from Ref.~\cite{Wang2012_ChinPhysC36-1603}.
$\beta_\mathrm{exp}$ represents experimental quadrupole deformations taken
from Ref.~\cite{Raman2001_ADNDT78-1}. 
The data with ``*'' stand for estimated values \cite{Wang2012_ChinPhysC36-1603}.
All energies are in MeV.
}
\begin{ruledtabular}
\begin{tabular}{lrrrrrrl}
 Nucleus   & $\beta_{20}$ & $\beta_{30}$ & $\beta_{32}$ & $\beta_{40}$ & $E_{\rm cal}$  & $\beta_{\rm exp}$ & $E_{\rm exp}$   \\ \hline
$^{100}$Zr & $0.44$      & $0.00$              & $0.00$             & $0.22$          & $-851.90$     & $0.36$                   & $-852.22$          \\
$^{102}$Zr & $-0.24$     & $0.00$              & $0.00$             & $0.04$          & $-863.10$     & $0.43$                   & $-863.57$          \\
$^{104}$Zr & $-0.24$     & $0.00$              & $0.00$             & $0.02$          & $-874.01$     & $0.47$                   & $-873.85$          \\
$^{106}$Zr & $-0.24$     & $0.00$              & $0.00$             & $0.01$          & $-884.18$     &                               & $-883.19^{*}$      \\
$^{108}$Zr & $-0.24$     & $0.00$              & $0.00$             & $0.00$         & $-893.16$      &                               & $-891.76^{*}$      \\
$^{110}$Zr & $0.00$      & $0.00$               & $0.18$             & $-0.04$        & $-902.46$      &                               & $-899.47^{*}$      \\
$^{112}$Zr & $0.00$      & $0.00$               & $0.13$             & $-0.02$        & $-908.79$      &                               & $-906.53^{*}$      \\
$^{114}$Zr & $-0.04$     & $0.13$              & $0.00$             & $0.02$          & $-915.50$       &                               &                             \\
\end{tabular}
\end{ruledtabular}
\end{table}

\begin{table}
\centering
\caption{\label{tab:mini_DDPC1} %
The binding energy (relative to the ground state) for various energy minima 
(when they exist)
in Zr isotopes obtained in the MDC-RHB calculations with the functional DD-PC1.
All energies are in MeV.
}
\begin{ruledtabular}
\begin{tabular}{lrrrrrrrr}
 Nucleus        & $^{100}$Zr & $^{102}$Zr & $^{104}$Zr & $^{106}$Zr & $^{108}$Zr & $^{110}$Zr & $^{112}$Zr  & $^{114}$Zr \\ \hline
 Oblate          & $0.21$        & $0.00$        & $0.00$       & $0.00$       & $0.00$       & $1.58$        & $0.44$        &  $0.19$       \\
 Pear-like       &                    &                   &  $2.72$      & $2.22$       & $1.12$       & $1.25$        & $0.08$        &  $0.00$       \\
 Tetrahedral   &                   &                     & $2.53$       & $1.85$       & $0.49$       & $0.00$        & $0.00$        &  $0.20$       \\
 Prolate          & $0.00$       & $0.10$        & $0.63$       & $0.89$       & $0.53$       & $2.55$        & $2.02$        & $4.02$       \\
\end{tabular}
\end{ruledtabular}
\end{table}

\begin{figure}
 \includegraphics[width=0.45\textwidth]{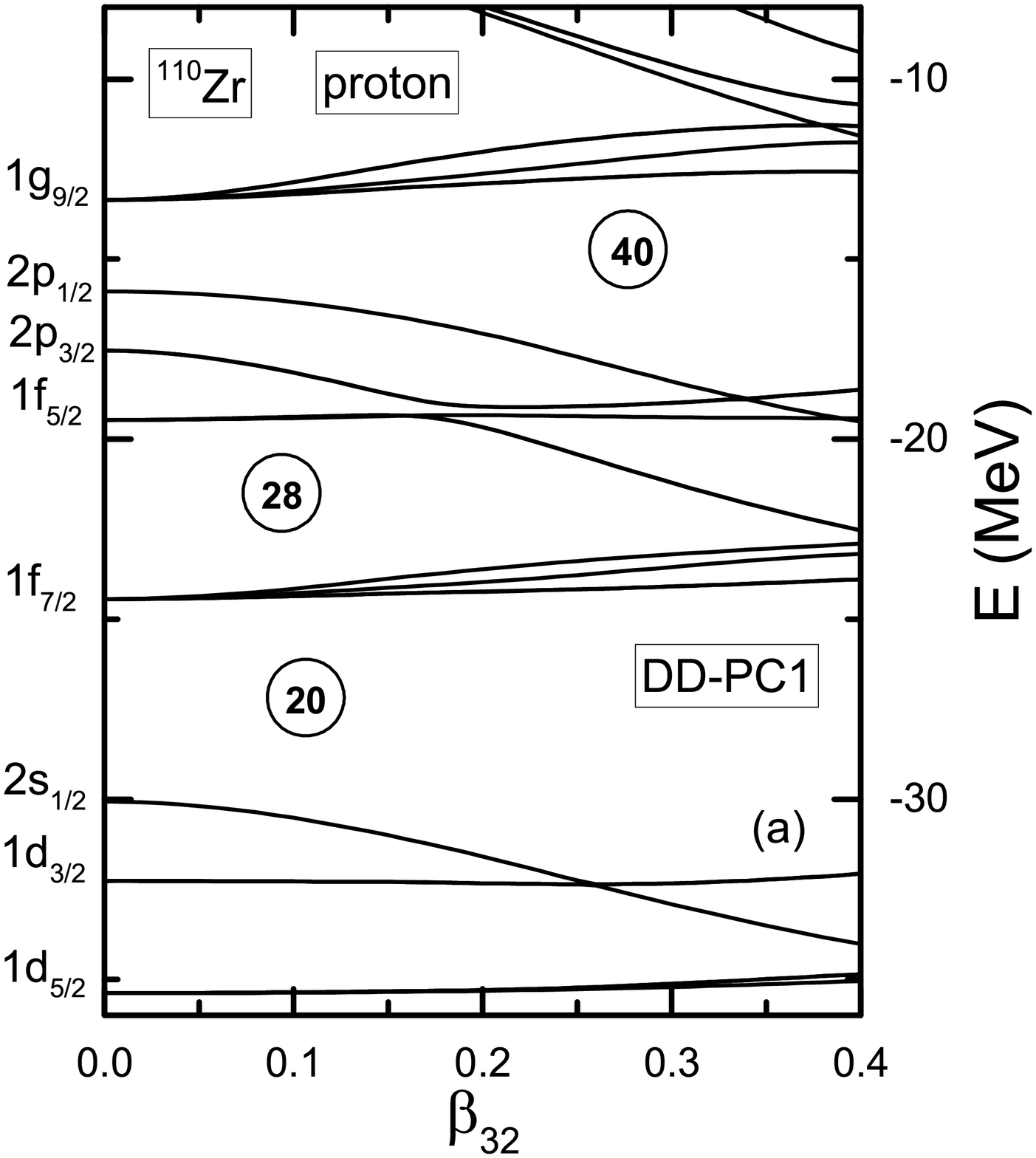}
 \includegraphics[width=0.45\textwidth]{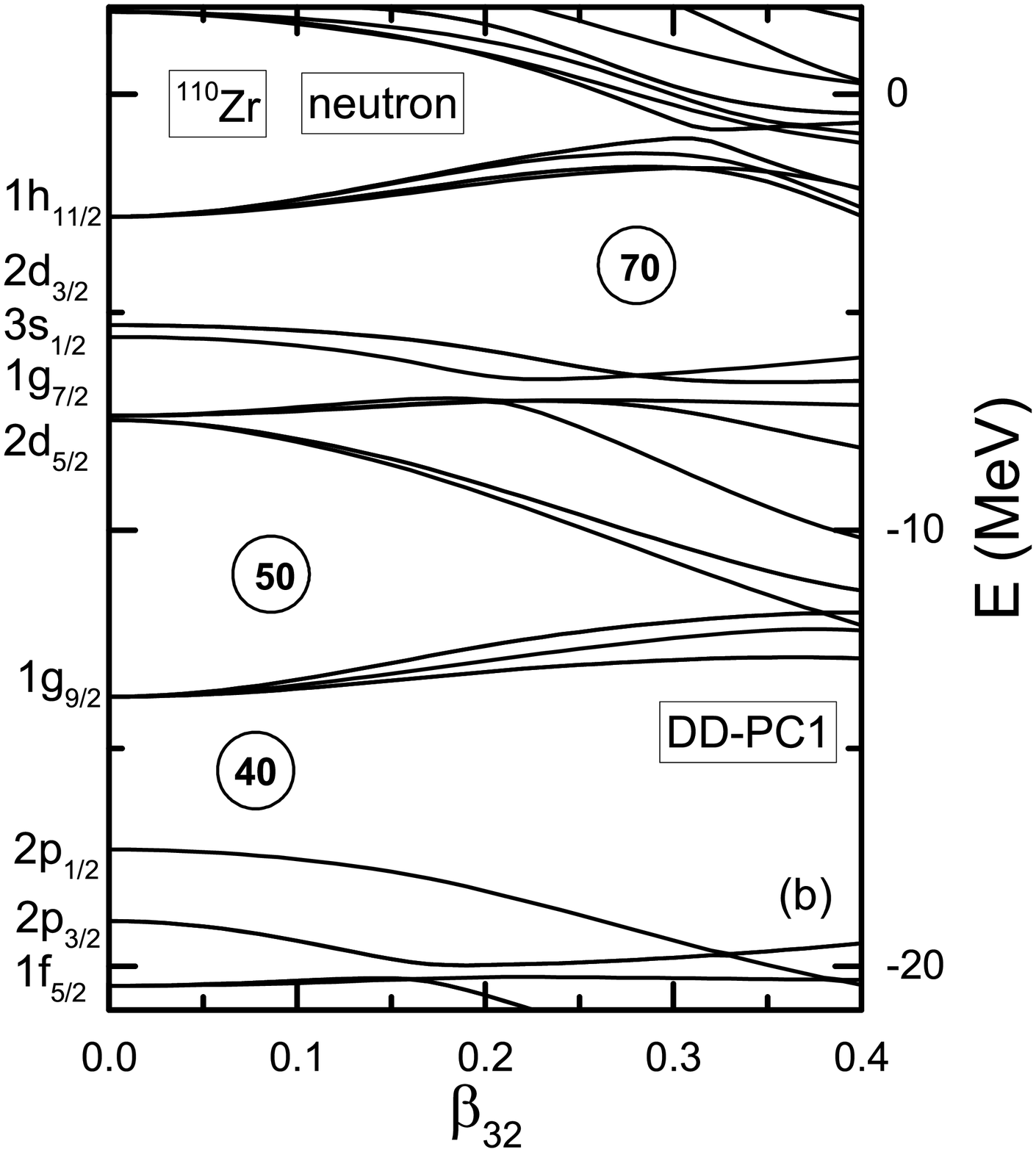}
\caption{\label{fig:lev} %
The single-particle levels near the Fermi surface for protons (a)
and neutrons (b) of $^{110}$Zr as functions of $\beta_{32}$ with $\beta_{20}$ fixed at zero.
The functional DD-PC1 is used in the MDC-RHB calculations.
}
\end{figure}

From Fig.~\ref{fig:Zr_DDPC1} and Table~\ref{tab:DDPC1_Gro}, 
one finds that the ground states of $^{100-108}$Zr are well deformed 
with large quadrupole deformation.  
For $^{108}$Zr, a tetrahedral isomeric state is also predicted.
The ground states of $^{110,112}$Zr are predicted to have tetrahedral shape 
and the most pronounced effect from the $\beta_{32}$ distortion is 
in $^{110}$Zr with the functional DD-PC1.
The formation of the tetrahedral ground state around $^{110}$Zr can 
be traced back to the large energy gaps at $Z=40$ 
and $N=70$ in the single-particle levels when $\beta_{32}$ deformation is included.
In Fig.~\ref{fig:lev}, we show the single-particle levels of $^{110}$Zr 
near the Fermi surface as functions of $\beta_{32}$ with $\beta_{20}$ fixed at zero.
Due to the tetrahedral symmetry, the single-particle levels split into 
multiplets with degeneracies equal to the irreducible representations 
of the $T^D_d$ group.
For example, the spherical levels with $j=1/2$ and $j=3/2$ are two-fold and 
four-fold degenerate and they can be reduced to the two-dimensional (2D) 
irreducible representation $E_1'$ and four-dimensional (4D) irreducible representation $G'$ 
of the $T^D_d$ group, respectively.
Such degeneracies are kept as $\beta_{32}$ increases.     
The spherical levels with $j=5/2$ can be reduced to the 2D irreducible 
representation $E_2'$ and 4D irreducible representation $G'$ of the $T^D_d$ group.
These levels split into two levels with degeneracies 2 and 4, respectively, 
as $\beta_{32}$ increases.
The reduction of spherical levels (up to $j=11/2$) to the irreducible 
representation of the $T^D_d$ group are listed in Table~\ref{tab:levels}.
For protons, as shown in Fig.~\ref{fig:lev}(a), 
the magic gap $Z\simeq 20$ is enhanced while the gap at $Z \simeq 28$ is suppressed 
as $\beta_{32}$ increases.
At $Z\simeq 40$ a large energy gap shows up with $\beta_{32}$ increasing.
From Fig.~\ref{fig:lev}(b) we can see that large energy gaps appears at $N \simeq 40$ 
and 70 while the spherical magic gap around $N=50$ is suppressed as $\beta_{32}$ increases.
Due to large energy gaps at $Z \simeq 40$ and $N \simeq 70$, a strong $\beta_{32}$ effect 
is expected for $^{110}$Zr and nearby nuclei.

\begin{table}
\centering
\caption{\label{tab:levels} %
The reductions of the spherical levels (up to $j=11/2$) to 
the irreducible representations of the $T^D_d$ group.
The two 2D irreducible representations of the $T^D_d$ group are denoted by $E_1'$ and $E_2'$,
the 4D irreducible representation of the $T^D_d$ group is denoted by $G'$.
}
\begin{ruledtabular}
\begin{tabular}{ll}
 $j$      & Irreducible representations of $T^D_d$ group  \\ \hline
 1/2      & $E_1'$                     \\
 3/2      & $G'$                         \\
 5/2      & $E_2'+G'$                \\
 7/2      & $E_1'+E_2'+G'$      \\
 9/2      & $E_1'+2G'$             \\
 11/2    & $E_1'+E_2'+2G'$    \\
\end{tabular}
\end{ruledtabular}
\end{table}

\begin{figure}
 \includegraphics[width=0.45\textwidth]{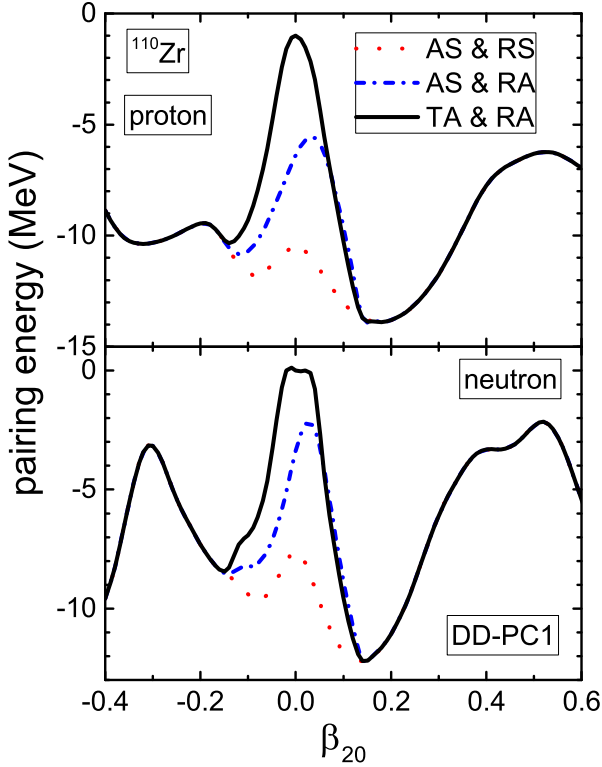}
\caption{(Color online)~\label{fig:pairing} %
The proton and neutron pairing energies of $^{110}$Zr as functions of $\beta_{20}$.
The dotted line denotes the results with the axial symmetry (AS) and 
reflection symmetry (RS) imposed,
the dash-dotted line denotes the results restricted to 
the AS and reflection asymmetry (RA),
while the solid line represents the results with both triaxial (TA)
and RA shapes allowed.
The functional DD-PC1 is used in the MDC-RHB calculations.
}
\end{figure}

In Fig.~\ref{fig:pairing} we show the proton and neutron pairing energies 
of $^{110}$Zr calculated with the functional DD-PC1.
The changes of the pairing energy as a function of $\beta_{20}$ are 
due to the variations of single-particle level
density near the Fermi surface when $\beta_{20}$ changes: 
lower level density near the Fermi surface usually leads to weaker pairing effects. 
When $\beta_{30}$ and $\beta_{32}$ deformations are not included, 
the weakest pairing effects are found around the prolate minimum 
(note that the three curves overlap around this minimum). 
However, the inclusion of $\beta_{30}$ or $\beta_{32}$ reduces 
the proton and neutron pairing effects around $\beta_{20}=0$ a lot.  
Especially for the case of $\beta_{32}$, both proton and neutron paring energies 
almost vanish around $\beta_{20}=0$, due to the large energy gap developed 
at $Z \simeq 40$ and $N \simeq 70$.

\begin{figure}
\centering
 \includegraphics[width=0.48\textwidth]{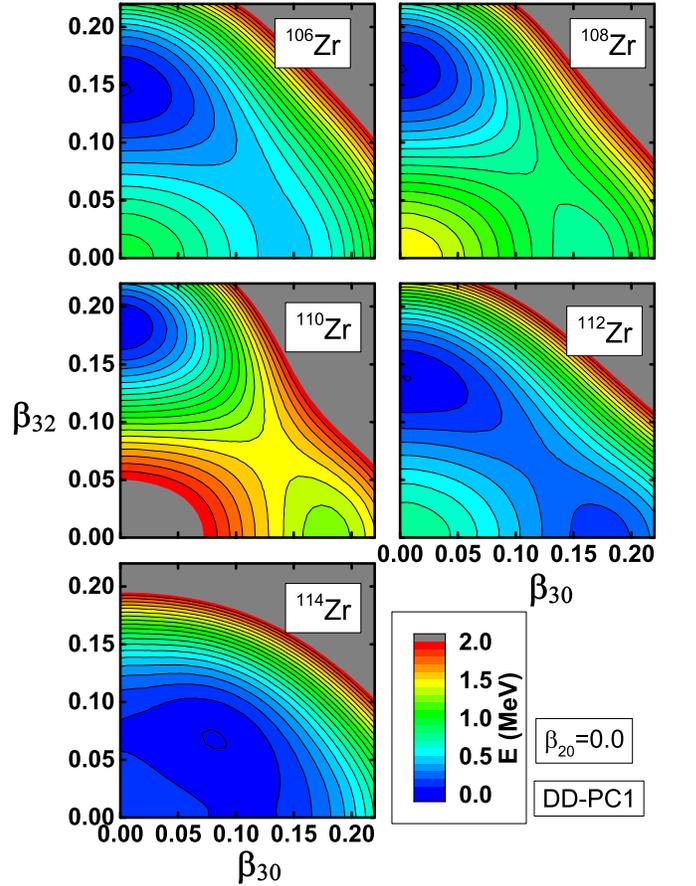}
\caption{\label{fig:DDPC1_Oct}(Color online) %
The potential energy surfaces of $^{106-114}$Zr in the $(\beta_{30}, \beta_{32})$ 
plane with $\beta_{20}$ fixed at zero. The contour interval is 0.1 MeV.
The functional DD-PC1 is used in the MDC-RHB calculations.
}
\end{figure}


From the above discussions, we know that the inclusion of $\beta_{30}$ and $\beta_{32}$ 
deformations can reduce the energies 
around $\beta_{20}=0$ for most of the neutron-rich Zr isotopes investigated here. 
As a result, with the functional DD-PC1, tetrahedral ground states 
and pear-like isomeric states are predicted for $^{110,112}$Zr 
and a pear-like ground state and a tetrahedral isomeric state are predicted for $^{114}$Zr.
To further investigate the properties and the relation between these two minima, 
we calculate the potential energy surfaces (PES) of $^{106-114}$Zr 
in the $(\beta_{30}, \beta_{32})$ plane with $\beta_{20}$ fixed at zero 
and show them in Fig.~\ref{fig:DDPC1_Oct}.
It is clearly seen that the pockets around tetrahedral minima 
are deeper than those around pear-like minima for $^{106-112}$Zr.
For example, the minimum with $\beta_{30} \approx 0.15$ of $^{110}$Zr 
is 1.25 MeV higher than the tetrahedral minimum with $\beta_{32} \approx 0.15$. 
From Fig.~\ref{fig:DDPC1_Oct}, one finds that the barriers 
separating the pear-like and the tetrahedral minima are very low.
For $^{106}$Zr, the barrier is almost invisible.
For $^{108}$Zr and $^{112}$Zr, the barrier heights are less than 0.2 MeV.
For $^{110}$Zr, the barrier is higher, but still less than 0.3 MeV.
In this sense, the pear-like isomeric states are rather unstable and
may hardly be observed for the nuclei discussed here.



We also preformed similar calculations with the functional PC-PK1.
By examining the obtained 1D potential energy curves and 2D potential
energy surfaces, it is found that with this functional and the same
pairing strength as used in calculations with the functional DD-PC1
(i.e., $G_n=G_0$, $G_p=1.12G_0$ with $G_0=728$ MeV$\cdot$fm$^3$), 
the tetrahedral and octupole deformations have less influence 
on Zr isotopes than they do with DD-PC1.
In this sense, the tetrahedral and octupole deformation effects are 
functional dependent.
With these two shape degrees of freedom considered, the potential energy
surfaces become softer around the local minima with $\beta_{20} \approx 0$.
The most pronounced effects from the tetrahedral or octupole distortions
happen in $^{110}$Zr for which a spherical shape is already obtained under
the assumption of axial and reflection symmetries. 
In the AS \& RA calculations, the ground state minimum for $^{110}$Zr
is lowered by about 0.1 MeV and in the TA \& RA calculations, the minimum
corresponding to the tetrahedral shape is lower by about 0.1 MeV
than that corresponds to an octupole shape. 

Experimentally, low-lying spectra for even-$A$ Zr isotopes have been established, 
from which it has been concluded that $^{100,102,104,106,108}$Zr are well deformed. 
However, there has been no such experimental information for $^{110}$Zr.
In some theoretical investigations (see, e.g., Ref.~\cite{Togashi2016_PRL117-172502})
it is predicted that $^{110}$Zr is also well deformed. 
In our calculations, 
$^{110}$Zr is predicted to be tetrahedral
because of the strong effects of shell closures around 
$Z \simeq 40$ and $N \simeq 70$.



\section{\label{sec:summary}Summary}

We have developed a multidimensionally constrained relativistic Hartree-Bogoliubov 
(MDC-RHB) model in which the pairing correlations are taken into account with 
the Bogoliubov transformation. 
In this model, the nuclear shape is assumed to be invariant under the reversion 
of $x$ and $y$ axes, thus, all shape degrees of freedom $\beta_{\lambda\mu}$ 
with even $\mu$ are included self-consistently. 
A separable pairing force of finite range is adopted in this model.
We solve the RHB equation in an axially deformed harmonic oscillator (ADHO) basis.
The convergence of the calculated results against the basis truncation is studied 
and it is shown that a reasonably large ADHO basis is able to provide a desired accuracy.

We have calculated the potential energy curves ($E\sim \beta_{20}$) of 
neutron-rich even-even Zr isotopes with the MDC-RHB model.
It is found that the $\beta_{32}$ deformation plays a very important role in 
the isomeric or ground states of these nuclei, especially for nuclei around $N=70$.
The ground state shape of $^{110}$Zr is predicted to be tetrahedral 
with both functionals DD-PC1 and PC-PK1.
$^{112}$Zr is also predicted to have a tetrahedral ground state with the functional DD-PC1.
We investigated the evolution of single-particle levels in $^{110}$Zr as a function
of the $\beta_{32}$ deformation. 
It is found that there are large energy gaps at $Z \simeq 40$ and $N \simeq 70$ which 
are responsible for the strong $\beta_{32}$ distortion effect.
The inclusion of $\beta_{30}$ deformation can also reduce 
the energy around $\beta_{20}=0$ and lead to minima with pear-like shapes for nuclei 
around $^{110}$Zr, but these minima are rather shallow.
By examining two-dimensional potential energy surfaces in the 
($\beta_{30}, \beta_{32}$) plane with $\beta_{20}$ fixed at zero, 
we found that the barrier separating the two minima with nonzero 
$\beta_{32}$ or nonzero $\beta_{30}$ is quite low.

\acknowledgments
This work has been supported by
the National Key Basic Research Program of China (Grant No. 2013CB834400),
the National Natural Science Foundation of China (Grants
No. 11120101005,
No. 11275248, 
No. 11475115, 
No. 11525524,
and
No. 11621131001),
and
the Knowledge Innovation Project of the Chinese Academy of Sciences (Grant No. KJCX2-EW-N01).
The computational results presented in this work have been obtained on
the High-performance Computing Cluster of SKLTP/ITP-CAS and
the ScGrid of the Supercomputing Center, Computer Network Information Center of
the Chinese Academy of Sciences.

\appendix

\section{Calculation of the pairing fields}
\label{app:PF}

The separable pairing force of finite range in the coordinate space reads
\begin{eqnarray}
 V(\bm{r}_{1}\sigma_{1},\bm{r}_{2}\sigma_{2},\bm{r}_{1}^{\prime}\sigma_{1}^{\prime},\bm{r}_{2}^{\prime}\sigma_{2}^{\prime})
&=& -G \delta(\bm{R}-\bm{R}^{\prime}) P(r)P(r^{\prime}) \nonumber \\
 &&\times \frac{1}{2} (1-P_{\sigma}),
\end{eqnarray}
where $\bm{R}=(\bm{r}_{1}+\bm{r}_{2}) / 2$ and $\bm{r}=\bm{r}_{1}-\bm{r}_{2}$
are the center-of-mass and relative coordinates, respectively. 
The corresponding matrix element in the axially deformed harmonic oscillator (ADHO) basis is
\begin{eqnarray}
V_{12,1^{\prime}2^{\prime}} 
& = & \sum_{\sigma_{1}\sigma_{2}}  \sum_{\sigma_{1}^{\prime}\sigma_{2}^{\prime}} 
      \int d^{3} \bm{r}_{1}d^{3}\bm{r}_{2}d^{3}\bm{r}_{1}^{\prime}d^{3}\bm{r}_{2}^{\prime} 
      \Phi_{1}^{*}(\bm{r}_{1}\sigma_{1})\Phi_{2}^{*}(\bm{r}_{2}\sigma_{2})  \nonumber \\
&  & \times V(\bm{r}_{1}\sigma_{1},\bm{r}_{2}\sigma_{2},\bm{r}_{1}^{\prime}\sigma_{1}^{\prime},\bm{r}_{2}^{\prime}\sigma_{2}^{\prime})
     \Phi_{1^{\prime}}(\bm{r}_{1}^{\prime}\sigma_{1}^{\prime})
     \Phi_{2^{\prime}}(\bm{r}_{2}^{\prime}\sigma_{2}). \nonumber \\
\end{eqnarray}
Referring to the structure of the basis wave functions Eq.~(\ref{eq:ppho}), 
$V_{12,1^{\prime}2^{\prime}}$ can be written as a product of integrals and sums,
\begin{equation}
V_{12,1^{\prime}2^{\prime}} = (C_{1}C_{2})^{*}(C_{1^{\prime}}C_{2^{\prime}})V_{12,1^{\prime}2^{\prime}}^{\sigma}
     V_{12,1^{\prime}2^{\prime}}^{z} V_{12,1^{\prime}2^{\prime}}^{\rho},
\end{equation}
with
\begin{eqnarray}
V_{12,1^{\prime}2^{\prime}}^{\sigma} 
& = & \sum_{\substack{\sigma_{1}\sigma_{2} \\ \sigma_{1}^{\prime}\sigma_{2}^{\prime}}} \chi_{s_{1}}(\sigma_{1}) \chi_{s_{2}}(\sigma_{2})  
       \frac{1}{2}(1-P_{\sigma}) \chi_{s_{1}^{\prime}}(\sigma_{1}^{\prime}) \chi_{s_{2}^{\prime}}(\sigma_{2}^{\prime}) \nonumber \\
& = & \frac{1}{2}(2s_{1}\delta_{\sigma_{1},\bar{\sigma_{2}}})(2s_{1}^{\prime}
    \delta_{\sigma_{1}^{\prime},\bar{\sigma_{2}^{\prime}}}),
\end{eqnarray}
\begin{equation}
\begin{aligned}
V_{12,1^{\prime}2^{\prime}}^{z} 
 = & \int dz_{1}dz_{2}dz_{1}^{\prime}dz_{2}^{\prime} \phi_{n_{z_{1}}}(z_{1}) \phi_{n_{z_{2}}}(z_{2}) \\
   &\times \left[\delta(Z-Z^{\prime}) P_{z}(z) P_{z}(z^{\prime})\right]
    \phi_{n_{z_{1}}^{\prime}}(z_{1}^{\prime}) \phi_{n_{z_{2}}^{\prime}}(z_{2}^{\prime}) \\
 = & \sqrt{2} \sum_{N_{z}}\left(\sum_{n}M_{N_{z}n}^{n_{z_{1}}n_{z_{2}}}A_{n}\right)
    \left(\sum_{n}M_{N_{z}n}^{n_{z_{1}}^{\prime}n_{z_{2}}^{\prime}}A_{n}\right), \\
\end{aligned}
\end{equation}
%
\begin{eqnarray}
V_{12,1^{\prime}2^{\prime}}^{\rho} 
 & = & \int d^{2}\bm{\rho}_{1}d^{2}\bm{\rho}_{2}d^{2}\bm{\rho}_{1}^{\prime}d^{2}\bm{\rho}_{2}^{\prime}  \nonumber \\
 &    &\times
     R_{n_{\rho_{1}}}^{m_{l_{1}}}(\rho_{1}) R_{n_{\rho_{2}}}^{m_{l_{2}}}(\rho_{2}) 
     \frac{1}{2\pi} e^{-i ( m_{l_{1}}\varphi_{1} + m_{l_{2}}\varphi_{2} )}  \nonumber \\
 &   &\times
     \left[\delta(\bm{\rho}-\bm{\rho}^{\prime})P_{\rho}(\rho)P_{\rho}(\rho^{\prime})\right]  \nonumber \\
 &   & \times 
    R_{n_{\rho_{1}}^{\prime}}^{m_{l_{1}}^{\prime}}(\rho_{1}) R_{n_{\rho_{2}}^{\prime}}^{m_{l_{2}}^{\prime}}(\rho_{2})
     \frac{1}{2\pi} e^{i ( m_{l_{1}}^{\prime} \varphi_{1} + m_{l_{2}}^{\prime} \varphi_{2} )} \nonumber \\
& = & \sqrt{2} \sum_{N_{\rho}M_{\rho}} 
    \left(\sum_{n}M_{N_{\rho}M_{\rho}n0}^{n_{\rho_{1}}m_{l_{1}}n_{\rho_{2}}m_{l_{2}}}B_{n}\right) \nonumber \\
&    & \times 
    \left(\sum_{n}M_{N_{\rho}M_{\rho}n0}^{n_{\rho_{1}}^{\prime}m_{l_{1}}^{\prime}
     n_{\rho_{2}}^{\prime}m_{l_{2}}^{\prime}}B_{n} \right),
\end{eqnarray}
where
\begin{eqnarray}
A_{n} & = & \int_{-\infty}^{\infty}\phi_{n}(z)P_{z}(\sqrt{2}z)dz, \nonumber \\
B_{n} & = & \sqrt{2\pi}\int_{0}^{\infty}P_{\rho}(\sqrt{2}\rho)R_{n}^{0}(\rho)\rho d\rho,
\end{eqnarray}
are constants. All these terms are in separable forms; 
so is the matrix element
\begin{equation}
V_{12,1^{\prime}2^{\prime}} = \sum_{N_{z}N_{\rho}M_{\rho}} 
      \left( W_{12}^{N_{z}N_{\rho} M_{\rho}} \right)^* 
             W_{1^{\prime}2^{\prime}}^{N_{z}N_{\rho}M_{\rho}},
\end{equation}
where
\begin{eqnarray}
W_{12}^{N_{z}N_{\rho}M_{\rho}} &=& (C_{1}C_{2})(2s_{1} \delta_{\sigma_{1},\bar{\sigma_{2}}})
    \left(\sum_{n}M_{N_{z}n}^{n_{z_{1}}n_{z_{2}}}A_{n}\right) \nonumber \\ 
 &&\times
    \left(\sum_{n}M_{N_{\rho}M_{\rho}n0}^{n_{\rho_{1}}m_{l_{1}}n_{\rho_{2}}m_{l_{2}}}B_{n}\right).
\end{eqnarray}
The Talmi-Moshinski brackets are given in the next appendix.

\section{The Talmi-Moshinski brackets}
\label{app:TM}

As usual the Talmi-Moshinski brackets are calculated by the generating
function method. Here we only show the final results and the details
can be found in Ref.~\cite{Chaos-Cador2004_IJQChem97-844}.

The one-dimensional Talmi-Moshinski bracket reads
\begin{eqnarray}
M_{N_{z}n_{z}}^{n_{z_{1}}n_{z_{2}}} 
& = & \int_{-\infty}^{\infty} dz_{1} dz_{2} \phi_{n_{z_{1}}}(z_{1}) 
        \phi_{n_{z_{2}}}(z_{2}) \phi_{N_{z}}(z_{+}) \phi_{n_{z}}(z_{-}) \nonumber \\
& = & \frac{1}{\sqrt{2^{N_{z}+n_{z}}}} \sqrt{\frac{n_{z_{1}}!n_{z_{2}}!}{N_{z}!n_{z}!}} \delta_{n_{z_{1}}+n_{z_{2}},N_{z}+n_{z}} \nonumber \\
&  & \times \sum_{s=0}^{n_{z}}\left(-1\right)^{s}
    \left(\begin{array}{c} N_{z} \\ n_{z_{1}}-n_{z}+s \end{array}\right)
    \left(\begin{array}{c} n_{z} \\ s \end{array}\right),
\end{eqnarray}
where $z_{+}=\frac{1}{\sqrt{2}}(z_{1}+z_{2})$ and $z_{-}=\frac{1}{\sqrt{2}}(z_{1}-z_{2})$.

The two-dimensional Talmi-Moshinski bracket is
\begin{widetext}
\begin{eqnarray}
M_{NMnm}^{n_{1}m_{1}n_{2}m_{2}} 
 &= & \int_{0}^{\infty}\rho_{1}d\rho_{1}\int_{0}^{2\pi}d\varphi_{1}
      	\int_{0}^{\infty}\rho_{2}d\rho_{2}\int_{0}^{2\pi}d\varphi_{2} 
    	R_{n_{1}}^{m_{1}}(\rho_{1}) e^{-im_{1}\varphi_{1}}R_{n_{2}}^{m_{2}}(\rho_{2}) e^{-im_{2}\varphi_{2}}
	R_{N}^{M}(\rho_{+})e^{iM\varphi_{+}}R_{n}^{m}(\rho_{-})e^{im\varphi_{-}} \nonumber \\
& = & \delta_{M+m,m_{1}+m_{2}} 
     \delta_{2n_{1}+|m_{1}|+2n_{2}+|m_{2}|,2N+|M|+2n+|m|} 
      \frac{(-1)^{n_{1}+n_{2}+N+n}}{\sqrt{2^{2n_{1}+|m_{1}|+2n_{2}+|m_{2}|}}}  
      \sqrt{\frac{N!(N+|M|)!n!(n+|m|)!}{n_{1}!(n_{1}+|m_{1}|)!n_{2}!(n_{2}+|m_{2}|)!}} \nonumber \\
 &  & \times
    \sum_{\substack{p_{1},q_{1},r_{1},s_{1} \\ p_{2},q_{2},r_{2},s_{2}}}^{n_{1} n_{2}} \sum_{t_{1} t_{2}}^{|m_{1}| |m_{2}|}
     (-1)^{r_{2}+s_{2}+|m_{2}|-t_{2}} \delta_{p_{2}r_{2}s_{2}t_{2}}^{p_{1}r_{1}s_{1}t_{1}} 
    \left(\begin{array}{c} |m_{1}| \\ t_{1} \end{array}\right) 
    \left(\begin{array}{c} |m_{2}| \\ t_{2} \end{array}\right)
    \left(\begin{array}{cccc} & n_{1} \\ p_{1} & q_{1} & r_{1} & s_{1} \end{array}\right) 
    \left(\begin{array}{cccc} & n_{2} \\ p_{2} & q_{2} & r_{2} & s_{2} \end{array}\right),\nonumber \\ 
\end{eqnarray}
where the variables $(\rho_{\pm},\varphi_{\pm})$ are defined as
\begin{eqnarray}
\rho_{\pm}\cos\varphi_{\pm} & = & \frac{1}{\sqrt{2}}(\rho_{1}\cos\varphi_{1}\pm\rho_{2}\cos\varphi_{2}),\nonumber \\
\rho_{\pm}\sin\varphi_{\pm} & = & \frac{1}{\sqrt{2}}(\rho_{1}\sin\varphi_{1}\pm\rho_{2}\sin\varphi_{2}).
\end{eqnarray}
The multinomial coefficients are
\begin{equation}
\left(\begin{array}{ccccc} &  & q \\ p_{1} & p_{2} & p_{3} & \cdots & p_{n} \end{array}\right)
= \delta_{\sum_{k}p_{k},q}\frac{q!}{\prod_{k=1}^{n}p_{k}},
\end{equation}
and the Kronecker $\delta$'s are 
\begin{equation}
\delta_{p_{2}r_{2}s_{2}t_{2}}^{p_{1}r_{1}s_{1}t_{1}}
=\left\{ \begin{array}{c}
    \delta_{t_{1}+t_{2},N-(p_{1}+s_{1})-(p_{2}+s_{2})+(|M|+M)/2}
    \delta_{0,N-(p_{1}+r_{1})-(p_{2}+r_{2})+\frac{1}{2}(|M|-M)},  \quad m_{1}m_{2} \geq 0 , \\
    \delta_{t_{1},N-(p_{1}+s_{1})-(p_{2}+s_{2})+(|M|+M)/2} 
    \delta_{t_{2},N-(p_{1}+r_{1})-(p_{2}+r_{2})+(|M|-M)/2},  \quad m_{1}m_{2} \leq 0.
\end{array}\right.
\end{equation}
\end{widetext}


%

\end{document}